\documentclass[usenatbib]{mnras}

\usepackage[utf8]{inputenc}
\usepackage[T1]{fontenc}
\usepackage{ae,aecompl}
\usepackage{geometry}
\usepackage{pdflscape}

\usepackage{graphicx}
\usepackage{float}
\usepackage{amssymb}
\usepackage{amsmath}
\usepackage{relsize}

\graphicspath{{Figures/}}

\def\beq{\begin{eqnarray}}
\def\eeq{\end{eqnarray}}

\def\ngal{n_{\mathrm{gal}}}

\def\siglogM{\sigma_{\log M}}
\def\Mmin{M_{\mathrm{min}}}
\def\satfrac{ M_1 / M_{\mathrm{min}}}

\def\M0oM1{ \frac{ M_0 }{  M_1 }}
\def\textM0oM1{\ln \left( M_0 / M_1 \right)}
\def\siglnMc{\sigma_{\ln M_c}}

\def\Mpch{h^{-1} \, \mathrm{Mpc}}
\def\MpchCubed{\mathrm{h^{-3}} \, \mathrm{Mpc}^3}
\def\invMpchCubed{h^3\,\mathrm{Mpc^{-3}}}
\def\Msun{\mathrm{M_{\odot}} \; h^{-1}}
\def\pimax{\Pi_{\mathrm{max}}}

\def\Sigcrit{\Sigma_{\mathrm{crit}}}

\def\Acon{A_\mathrm{con}}
\def\Qcen{Q_\mathrm{cen}}
\def\Qsat{Q_\mathrm{sat}}
\def\Alens{A_\mathrm{lens}}

\def\fcen{f_\mathrm{cen}}
\def\cvir{c_\mathrm{vir}}
\def\fcen{f_\mathrm{cen}}
\def\nfid{n_\mathrm{fid}}

\title[Galaxy-Galaxy Lensing and Clustering Forecasts for DES-Y6]{Exploiting Non-linear Scales in Galaxy-Galaxy Lensing and Galaxy Clustering: A Forecast for the Dark Energy Survey}

\author[A. N. Salcedo et al.]{Andr\'{e}s N. Salcedo$^{1}$\thanks{E-mail: salcedo.11@osu.edu},
David H. Weinberg$^{1,2}$,
Hao-Yi Wu$^{3}$  
and Benjamin D. Wibking$^{4}$
\\
$^{1}$ Department of Astronomy and Center for Cosmology and AstroParticle Physics, The Ohio State University, Columbus, OH 43210, USA \\
$^{2}$ Institute for Advanced Study, Princeton, NJ 08540, USA \\
$^{3}$ Department of Physics, Boise State University, Boise, ID 83725, USA \\
$^{4}$ Australian National University, Mount Stromlo Observatory, Cotter Road, Weston Creek, ACT 2611, Australia
}

\date{Accepted XXX. Received YYY; in original form ZZZ}

\pubyear{2021}

\begin{document}
\label{firstpage}
\pagerange{\pageref{firstpage}--\pageref{lastpage}}
\maketitle

\begin{abstract}
The combination of galaxy-galaxy lensing (GGL) and galaxy clustering is a powerful probe of low redshift matter clustering, especially if it is extended to the non-linear regime.  To this end, we extend the N-body and halo occupation distribution (HOD) emulator method of \citet{Wibking_et_al_2020} to model the redMaGiC sample of colour-selected passive galaxies in the Dark Energy Survey (DES), adding parameters that describe central galaxy incompleteness, galaxy assembly bias, and a scale-independent multiplicative lensing bias $\Alens$.  We use this emulator to forecast cosmological constraints attainable from the GGL surface density profile $\Delta\Sigma(r_p)$ and the projected galaxy correlation function $w_{p,gg}(r_p)$ in the final (Year 6) DES data set over scales $r_p=0.3-30.0\, \Mpch$.  For a $3\%$ prior on $\Alens$ we forecast precisions of $1.9\%$, $2.0\%$, and $1.9\%$ on $\Omega_m$, $\sigma_8$, and $S_8 \equiv \sigma_8\Omega_m^{0.5}$, marginalized over all halo occupation distribution (HOD) parameters as well as $\Alens$ and a point-mass contribution to $\Delta\Sigma$.  Adding scales $r_p=0.3-3.0\, \Mpch$ improves the $S_8$ precision by a factor of ${\sim}1.6$ relative to a large scale ($3.0-30.0\, \Mpch$) analysis, equivalent to increasing the survey area by a factor of ${\sim}2.6$.  Sharpening the $\Alens$ prior to $1\%$ further improves the $S_8$ precision by a factor of $1.7$ (to $1.1\%$), and it amplifies the gain from including non-linear scales. Our emulator achieves percent-level accuracy similar to the projected DES statistical uncertainties, demonstrating the feasibility of a fully non-linear analysis.  Obtaining precise parameter constraints from multiple galaxy types and from measurements that span linear and non-linear clustering offers many opportunities for internal cross-checks, which can diagnose systematics and demonstrate the robustness of cosmological results.
\end{abstract}


\section{Introduction}

Understanding the origin of cosmic acceleration remains the most pressing
challenge of contemporary cosmology.  Ambitious cosmological surveys are using a variety of observational probes to measure the histories of cosmic expansion and the growth of matter clustering with high precision over a wide span of redshift \citep[for reviews see e.g.,][]{Frieman_AnnRev_et_al_2008, Weinberg_PhR_2013}. Comparing expansion history and structure growth is critical to testing whether cosmic acceleration reflects a breakdown of General Relativity (GR) on cosmological scales or a form of dark energy that exerts repulsive gravity within GR.  With present data sets, the most powerful constraints on low redshift matter clustering come from large area weak lensing surveys, which can measure matter clustering directly through cosmic shear or by combining galaxy-galaxy lensing (GGL) with galaxy clustering.

This paper presents methodology for and forecasts of the precision obtainable with the combination of GGL and galaxy clustering in the final data sets from the Dark Energy Survey \citep[DES;][]{DES_2005,DES_3x2pt_2021}, building on the work of \citet{Wibking_et_al_2019, Wibking_et_al_2020}. GGL measures correlations between foreground lens galaxies and a shear map of background source galaxies to infer the lens galaxies' mean excess surface density profile $\Delta\Sigma(r_p)$, which is proportional to the product of the  matter density parameter $\Omega_m$ and the galaxy-matter cross-correlation function $\xi_{gm}$.  On scales large enough to be described by linear perturbation theory one expects $\xi_{gm}=b_g\xi_{mm}$ and $\xi_{gg}=b_g^2\xi_{mm}$, where $b_g$ is the galaxy bias factor and $\xi_{gg}$ and $\xi_{mm}$ are the galaxy and matter autocorrelation functions, respectively.  One can therefore combine GGL and $\xi_{gg}$ to cancel the unknown $b_g$ and constrain $\Omega_m\sqrt{\xi_{mm}} \propto \Omega_m\sigma_8$, where $\sigma_8$, the RMS linear theory matter overdensity fluctuation in spheres of radius $8.0 \, \Mpch$ at $z=0$, is an overall scaling of the amplitude of matter  fluctuations.  In practice the best constrained parameter combination is closer to $S_8 \equiv \sigma_8\Omega_m^{0.5}$.

Interpreting GGL and clustering measurements on smaller scales requires a model for the relation between galaxies and dark matter in the non-linear regime, such as the halo occupation distribution \citep[HOD;][]{Jing_Mo_Borner_1998,Peacock_Smith_2000,Scoccimarro_et_al_2001, Berlind_2002} or sub-halo abundance matching \citep{Conroy_et_al_2006,Vale_Ostriker_2006}. Although these models require additional free parameters, non-linear clustering data can constrain them, so extending to the non-linear regime of GGL and $\xi_{gg}$ can potentially achieve much tighter constraints on cosmological parameters \citep{Yoo_et_al_2006,Zheng_Weinberg_2007, Cacciato_et_al_2009,Cacciato_et_al_2012,Cacciato_et_al_2013, Leauthaud_et_al_2011,Yoo_Seljak_2012,More_et_al_2013}. The stakes of this effort are illustrated by a number of recent studies finding that the amplitude of matter clustering inferred from GGL + galaxy clustering is 5-10\% lower than the amplitude predicted by extrapolating CMB anisotropies forward to low redshift assuming a $\Lambda$CDM cosmological model \citep{Mandelbaum_2013,More_et_al_2015, Leauthaud_et_al_2017,Joudaki_KIDS_et_al_2018, Singh_et_al_2020,Wibking_et_al_2020,Krolewski_et_al_2021}.\footnote{We use $\Lambda$CDM to denote a model with inflationary primordial fluctuations, cold dark matter, a cosmological constant, and a flat universe.} The conflict is strongest on non-linear scales, where the measurement precision is highest but the demands on the accuracy of non-linear modeling are the most stringent \citep{Leauthaud_et_al_2017,Lange_et_al_2019, Lange_et_al_2021}. Many cosmic shear studies also find clustering amplitudes lower than the CMB-based prediction, but the discrepancy is less statistically significant \citep{Jee_et_al_2016,Hildebrandt_KIDS_et_al_2017,Hikage_HSC_et_al_2019,DES_shearY3_et_al_2021, Secco_DES_et_al_2021}. The recent ``3x2pt'' analysis of the Year 3 (Y3) DES data, which combines cosmic shear, GGL, and galaxy clustering on scales adequately described by linear theory, yields results compatible with CMB-based $\Lambda$CDM predictions, but also compatible with the lower amplitudes reported in the studies listed above \citep{DES_3x2pt_2021}.

In this study we adopt the HOD framework to model non-linear galaxy bias. HOD methods statistically specify the relationship between galaxies and their host haloes, primarily as a function of host halo mass.  In a cosmological context the most important question is whether or not a given HOD parameterization is flexible enough to model the non-linear galaxy bias without producing biased cosmological constraints.  One of the most important sources of systematic uncertainty in the galaxy-halo connection is the potential presence of galaxy assembly bias. Galaxy assembly bias refers to the potential for the galaxy occupation inside haloes of the same mass to vary with respect to a secondary halo property.  In combination with {\emph{halo}} assembly bias \citep[e.g.][]{Sheth_Tormen_2004, Gao_2005, Harker_et_al_2006, Wechsler_2006, Gao_White_2007, Jing2007, Wang_Mo_Jing_2007, Li_Mo_Gao_2008,Faltenbacher_White_2010, Mao_Zentner_Wechsler_2018, Salcedo_2018, Sato-Polito_et_al_2018, Xu_Zheng_2018, Johnson_et_al_2019}, the potential for halo clustering at fixed mass to vary with respect to a secondary halo property, this can modify
the large scale galaxy clustering making predictions from a standard HOD model inaccurate \citep{Croton_et_al_2007, Zu_et_al_2008, McCarthy_et_al_2019}.  To provide our HOD framework the flexibility to describe potential galaxy assembly bias we adopt the modifications of \citet{Salcedo_et_al_2020b} \citep[see also][]{Wibking_et_al_2019, McEwen_2018, Salcedo_et_al_2020, Xu_et_al_2020}.  For our forecasts, we focus on the DES redMaGiC galaxy sample \citep{Rozo_Rykoff_redmagic_et_al_2016}, which uses colour selection to identify passive galaxies that allow precise photometric redshifts. To model a colour-selected sample, we also extend the usual HOD formulation to include a parameter that allows for ``central galaxy incompleteness,'' i.e., for high mass halos that do not contain central galaxies passing the sample's colour cuts.

We adopt this HOD framework to populate N-body simulations from the
AbacusCosmos suite \citep{Garrison_et_al_2018} in order to model the
GGL excess surface density $\Delta \Sigma (r_p)$ and projected
galaxy correlation function $w_{p,gg} (r_p)$ on scales $0.3 \, \Mpch < r_p < 30.0 \, \Mpch$ as a function of HOD and $w\mathrm{CDM}$ cosmological parameters.  To accurately model this datavector down to small scales we adopt and extend the Gaussian process emulation scheme of \citet{Wibking_et_al_2020}.  This emulation is done over a large HOD and cosmological parameter space centered on a fiducial model that roughly describes the high-density DES redMaGiC sample.  We use this emulator to compute derivatives of $\Delta \Sigma$ and $w_{p,gg}$ with respect to HOD and cosmological parameters, which we then use to forecast a cosmological analysis of DES redMaGiC GGL and clustering.  We devote particular attention to the importance of the small scales in such an analysis and also to the ability of the datavector to break the degeneracy between cosmology and systematic uncertainties in lensing calibration.  The technical development behind producing our forecasts is aimed at enabling a fully non-linear GGL and clustering analysis of the final DES data release, which we predict to yield percent-level constraints on the amplitude of matter clustering.

Our forecasts could prove optimistic if observational or theoretical systematics in the final DES data turn out to be larger than we have assumed.  For example, the Y3 cosmology analysis identifies systematics in the clustering measurements of the redMaGiC sample \citep{DES_3x2pt_2021}, while we have assumed that systematic uncertainties in $w_{p,gg}$ will be negligible. Nonetheless, our forecasts play a valuable role in demonstrating what DES GGL+clustering should be able to achieve if systematics are well controlled, thus also demonstrating the level of systematics control that is required. A key finding of our analysis is that modeling GGL+clustering into non-linear scales can achieve gains in cosmological parameter precision that are equivalent factors of $2.5-8.0$ increases in survey area, e.g., to the difference between a 5-year weak lensing survey and a survey lasting one to several decades. Doing the additional work needed to realize these gains is a promising investment.

The next section describes our numerical simulations and HOD modeling
methodology.  Section \ref{sec:emu} defines our clustering and lensing
statistics, then describes our emulation methodology and derives the sensitivity of our datavector to HOD and cosmological parameters.  Section \ref{sec:cov} describes how we compute covariance matrices for our Fisher forecasts, based on expectations for the final DES data release.  In section \ref{sec:forecasts} we present our main forecast results, which combine the derivatives computed in section \ref{sec:deriv} with the covariance matrices of section \ref{sec:cov} to derive constraints on $\Omega_m$, $\sigma_8$, and $S_8$.  We summarise our results and conclude in section \ref{sec:conc}.

\section{Constructing Mock Galaxy Catalogs}

\subsection{Simulations and Halo Identification}
\label{sec:sims}

\begin{table*}
   \centering
   \caption{Fiducial Model Parameters (HOD and Cosmological).}
    \begin{tabular}{cccl}      
    \hline
      Parameter  & Fiducial Value & Sampling Range & Description \\
 	\hline
 	$\ngal \times 10^3$ & $1.0 \; \invMpchCubed$ & $[0.8, 1.2] \times \; \invMpchCubed$ & galaxy number density\\
 	$\siglogM$ & $0.6$ & $[0.4, 0.8]$& width of central occupation cutoff\\
 	$\frac{M_1}{M_\mathrm{min}}$ & $30.0$ & $[20.0, 50.0]$ & satellite fraction parameter\\
 	$\M0oM1$ & $0.2$ & $-$ & satellite cutoff parameter\\
 	$\alpha$ & $1.5$ & $[1.2,1.8]$ & slope of satellite occupation power law\\
 	\hline
 	$\fcen$ & $0.6$ & $[0.4, 0.8]$ & central incompleteness factor\\
 	$\Acon$ & $1.0$ & $[0.5, 2.0]$ & galaxy concentration factor\\
 	$\Qcen$ & $0.0$ & $[-0.3, 0.3]$ & central galaxy assembly bias parameter\\
 	$\Qsat$ & $0.0$ & $[-0.3, 0.3]$ & satellite galaxy assembly bias parameter\\
 	\hline
 	$\Alens$ & $1.0$ & $-$ & scale independent lensing bias parameter\\
	\hline
	$\Omega_m$ & $0.314$ &  $[0.253, 0.367]$ & cosmological matter density\\
	$\sigma_8$ & $0.83$ & $[0.65, 1.0]$ & power spectrum amplitude\\
	$H_0$ & $67.26$ & $[61.567, 74.793]$ & Hubble constant\\
	$w_0$ & $-1.00$ & $[-1.370, -0.655]$ & equation of state of dark energy\\
	$n_s$ & $0.9652$ & $[0.9300, 0.9898]$ & scalar spectral index\\
	\hline
   \end{tabular}
\label{table:params}
\end{table*}

We use 40 AbacusCosmos simulations in our analysis \citep{Garrison_et_al_2018}. These simulations are run with a variety of $w\mathrm{CDM}$ cosmologies centered on the \citet{Planck_2016} cosmology with fixed phases. The 40 cosmologies are selected using a Latin hypercube method \citep{Heitmann_et_al_2009} optimized to maximize the distance between points. These cosmologies are sampled from a parameter space consisting of the union of CMB, BAO and SN results described in \cite{AndersonBOSS_et_al_2014}. We utilize the larger $1100.0 \; \MpchCubed$ set of boxes with mass resolution of $10^{10} \; \Msun$. 

Haloes were identified from particle snapshots using the software package {\sc{rockstar}} version 0.99.9-RC3+ \citep{Behroozi_2013}. We use strict (i.e., without unbinding) spherical overdensity (SO) halo masses around the halo centres identified by {\sc{rockstar}}, rather than the default phase-space FOF-like masses output by {\sc{rockstar}}. For finding haloes {\sc{rockstar}} uses a primary definition set to the virial mass of \citet{Bryan_1998}. However, after identification, we adopt the $M_{200b}$ mass definition, i.e.,  the mass enclosed by a spherical overdensity of 200 times the mean matter density at a given redshift and cosmology. Distinct haloes identified with the $M_{vir}$ definition are not reclassified as subhalos under the $M_{200b}$ definition; such reclassification would affect a negligible fraction of halos. We identify halos above 20 particles, and we only use distinct halos (not subhalos) when creating galaxy populations.

\subsection{HOD Modeling}

Similar to our previous papers we populate simulated haloes with galaxies according to a halo occupation distribution (HOD) framework \citep[e.g.][]{Jing_Mo_Borner_1998, Benson_et_al_2000, Ma_Fry_2000,Peacock_Smith_2000, Seljak_2000, Scoccimarro_et_al_2001, Berlind_2002, Cooray_Sheth_2002, Yang_et_al_2003, vdBosch_et_al_2003b, Zheng_et_al_2005, Cooray_2006, Mandelbaum_et_al_2006, Zheng_2009, Zehavi_et_al_2011, Coupon_et_al_2012, Leauthaud_et_al_2012, Guo_2014, Zu_Mandelbaum_12-2015, Zehavi_et_al_2018}. We extend this framework to include central incompleteness, galaxy assembly bias, and the possibility for the galaxy profile to deviate from that of its host's matter profile. We parametrize the mean central and satellite occupations of our haloes with a modified form of the widely used equations \citep{Zheng_et_al_2005}
\begin{align}
\left \langle N_{\mathrm{cen}}(M_h) \right \rangle &= \frac{\fcen}{2} \left [ 1 + \mathrm{erf} \left ( \frac{\log M_h - \log M_{\mathrm{min}}}{\siglogM}\right ) \right ], \label{eq:cen_HOD} \\ 
\left \langle N_{\mathrm{sat}}(M_h) \right \rangle &= \frac{\left \langle N_{\mathrm{cen}} (M_h) \right \rangle}{\fcen} \left ( \frac{M_h - M_0}{M_1} \right )^{\alpha}. \label{eq:sat_HOD}
\end{align}
The new parameter $\fcen$ allows only a fraction of high mass haloes to contain central galaxies that satisfy the sample selection criteria. Incompleteness may be present in any galaxy sample, but is particularly important for us to model because we are forecasting for an analysis that utilizes redMaGiC \citep{Rozo_Rykoff_redmagic_et_al_2016} selected galaxies. These galaxies are known to exhibit central incompleteness because of the strict colour cuts applied in their selection. The fraction of satellite galaxies that pass selection criteria is already encoded within the parameter $M_1$.

The actual numbers of centrals and satellites placed into each halo is drawn randomly from binomial and Poisson distributions, respectively, with the mean occupations given above. Centrals are placed at the center of their host halo, while satellites are distributed according to a Navarro-Frenk-White profile \citep[NFW;][]{NFW_1997},
\beq
\rho_\mathrm{gal}(r) = \rho_m (r | \Acon \times \cvir)
\eeq
parameterized by halo concentration $\cvir{=}r_h/r_s$ with the parameter $\Acon$ included to allow for the galaxy profile to deviate from that of the matter. As in \citet{Salcedo_et_al_2020}, we use the fits of \citet{Correa_2015} to assign halo concentrations because they were calibrated using significantly higher resolution simulations than our AbacusCosmos boxes.

Following \citet{Wibking_et_al_2019} \citep[see also][]{McEwen_2018,Salcedo_et_al_2020,Xu_et_al_2020} we allow for the possibility of galaxy assembly bias. Galaxy assembly bias refers to the possibility for galaxy occupation at fixed host halo mass to depend on properties other than halo mass. In combination with {\emph{halo}} assembly bias this can boost the large scale clustering of galaxies \citep[e.g.][]{Croton_et_al_2007,Zu_et_al_2008}, and it represents an important source of systematic uncertainty in current cosmological analyses. It is currently unclear which halo internal property, if any, is responsible for galaxy assembly bias. However, in the context of a cosmological analysis, in which assembly bias is treated as a nuisance effect to be marginalized over, it is only important to {\emph{characterize}} its potential effects. Therefore we choose to allow the central and satellite occupations to vary on a halo-by-halo basis based on the matter overdensity measured in a top-hat spheres of radius $8.0 \, \Mpch$ centered on each individual halo $\delta^m_8$. This environmental dependence is written as
\begin{align}
    \log M_\mathrm{min} &= \log M_\mathrm{min,0} + \Qcen (\tilde{\delta}^m_8 - 0.5),\\
    \log M_1 &= \log M_\mathrm{1,0} + \Qsat (\tilde{\delta}^m_8 - 0.5),
\end{align}
where $\Qcen$ and $\Qsat$ express the strength of the dependence of $M_\mathrm{min}$ and $M_1$ respectively on environment and $\tilde{\delta}^m_8 \in [0,1]$ is the normalized rank of $\delta^m_8$ within a narrow mass bin. In this parametrization the case of $\Qcen{=}\Qsat{=}0.0$ corresponds to having no assembly bias. This parameterisation has been found to provide a reasonable description of galaxy assembly bias effects in semi-analytic models and hydrodynamic simulations \citep[e.g.][]{Artale_et_al_2018, Zehavi_et_al_2018, Bose_et_al_2019, Contreras_et_al_2019}.

The Y3 DES $3{\times}2$pt. cosmological analysis \citep{DES_3x2pt_2021} considered both the redMaGiC galaxy sample and an apparent magnitude limited sample \citep{Porredon_MAGLIM_DES_et_al_2021}, adopting the latter for its fiducial results. We expect that our HOD parameterization would adequately represent this magnitude limited sample, but the fiducial parameters would be quite different from those for redMaGiC, with higher $\ngal$, higher $\fcen$, shallower $\alpha$, and perhaps smaller $\siglogM$, based on SDSS results at low redshift \citep{Zehavi_et_al_2011}. Because of the higher $\ngal$, the magnitude limited sample should yield smaller statistical errors, particularly for $\Delta \Sigma$, and might therefore achieve tighter statistical constraints than those forecast here. However, there are additional complications in modeling this sample because of the lower precision of photometric redshifts, and we have not investigated the impact of these complications.

\section{Emulation of Cosmological Observables}
\label{sec:emu}

\subsection{Clustering and Weak-Lensing Statistics}
\label{sec:obs}

We use {\sc{corrfunc}} \citep{Sinha_2017} to compute the real-space galaxy autocorrelation function $\xi_{gg}(r_p, \pi)$ and galaxy-matter cross-correlation function $\xi_{gm}(r_p, \pi)$ in 20 equal logarithmically spaced bins of $r_p$ covering scales $0.3 < r_p < 30.0 \; \Mpch$ and 100 equal linearly spaced bins out to $\pimax = 100.0 \; \Mpch$. These real-space correlation functions are used to calculate the more observationally motivated quantities $w_{p,gg}(r_p)$ and $\Delta \Sigma (r_p)$,
\beq
w_{p,AB}(r_p) = 2 \int_0^{\pimax} \xi_{AB}(r_p, \pi),
\eeq
\beq
\Delta \Sigma (r_p) = \Omega_m \rho_\mathrm{crit} \left[ \frac{2}{r^2_p} \int_0^{r_p} r' w_{p,gm}(r') d r' - w_{p,gm}(r_p)\right].
\eeq
For a given source redshift distribution, $\Delta \Sigma (r_p)$ is proportional to the observable tangential shear profile,
\beq
\gamma_t(r_p) = \frac{\Delta \Sigma (r_p)}{ \Sigcrit},
\eeq
where the  critical surface density $\Sigcrit$ is
\beq
\Sigcrit = \frac{c^2}{4 \pi G} \frac{H(z_\mathrm{src} - z_\mathrm{lens})D_c(z_\mathrm{src})}{D_c(z_\mathrm{lens}) \left[ D_c(z_\mathrm{src}) - D_c(z_\mathrm{lens}) \right] (1 + z_\mathrm{lens})},
\eeq
and where $D_c(z)$ denotes the comoving distance to redshift $z$ with the Heaviside step function $H$ enforcing the convention that $\Sigcrit(z_\mathrm{src} < z_\mathrm{lens})=0$. 

Errors in photometric redshift estimation can introduce errors into $\Sigcrit$ and therefore $\Delta \Sigma$. Additionally, errors in shear calibration will introduce errors in $\Delta \Sigma$ through $\gamma_t$. We characterize the effect of these errors by introducing a scale-independent lensing bias parameter $\Alens$, 
\beq
\Delta \Sigma_\mathrm{obs}(r_p) = \Alens \times \Delta \Sigma_\mathrm{true}(r_p).
\eeq
We include $\Alens$ as an additional nuisance parameter that we marginalize over in our forecasts in section \ref{sec:forecasts}. The DES Y3 analysis finds evidence of internal inconsistency bewteen the clustering and GGL of redMaGiC galaxies, which they tentatively ascribe to an undiagnosed systematic in the clustering measurements \citep{DES_3x2pt_2021, Pandey_DES_et_al_2021}. They model this effect with a nuisance parameter $X_\mathrm{lens}$ that scales the predicted GGL signal relative to clustering, inferring a value $X_\mathrm{lens} \approx 0.9$ rather than the theoretically expected $1.0$. We suspect that our forecasts would be similar if we replaced $\Alens$ with $X_\mathrm{lens}$ as a nuisance parameter {\emph{and}} adopted the same fractional prior ($3\%$ in our fiducial case). However, we have not investigated this alternative parameterization of systematics. Our forecasts implicitly assume that the systematics suggested by Y3 redMaGiC galaxy clustering will be controlled in the final analysis, at least to the level represented by our $\Alens$ prior.

In addition to a multiplicative lensing bias, so called ``boost'' factors will also affect the galaxy-galaxy lensing signal of redMaGiC galaxies. These are a correction to the measured lensing signal to account for the presence of lens-source clustering. In the case of redMaGiC galaxies, boost factors significantly impact the small scale lensing signal, but their uncertainties are relatively small \citep[e.g.][]{Prat_DES_et_al_2021} and subdominant to our statistical errors. Therefore we do not model boost factors for our forecast analysis.

Our choice to model the projected correlation function of redMaGiC selected galaxies is somewhat idiosyncratic, since these galaxies have photometrically estimated redshifts. Photometric samples are more commonly characterized by the angular correlation function in photo-z bins, whereas $w_{p,gg}$ uses the photo-z's of each pair of galaxies to estimate the separations of $r_p$ and $\pi$. The redMaGiC algorithm produces impressively precise photometric redshifts in the redshift range $z = 0.1 - 0.7$, roughly $1 - 2\%$ in terms of $1+z_\mathrm{phot}$ \citep{Rozo_Rykoff_redmagic_et_al_2016}. Given our fiducial cosmology, this precision corresponds to $30.0 - 60.0 \, \Mpch$ errors in line of sight distance. Because we integrate to $\pimax = 100.0 \, \Mpch$, the photo-z errors will mildly depress $w_{p,gg}(r_p)$ by a scale-independent factor \citep{WangZhaoyu_et_al_2019}. In this paper we ignore this effect, implicitly taking for granted our ability to model it with good enough knowledge of photo-z errors, and assuming its independence from HOD and cosmology. We examine this problem more fully in a forthcoming paper \citep{ZengC_et_al_prep}.

To model the dependence of $w_{p,gg}$ on our HOD and cosmological parameters we choose to directly emulate a halo-model correction \citep{Wibking_et_al_2020},
\beq
f_\mathrm{corr}(r_p) = \frac{w_{p,gg}^\mathrm{sim}(r_p)}{w_{p,gg}^\mathrm{model}(r_p)},
\eeq
where $w_{p,gg}^\mathrm{sim}$ is calculated using {\sc{corrfunc}} on our simulation mock galaxy catalogs, and $w_{p,gg}^\mathrm{model}$ is analytically calculated. This procedure has two major advantages, the first being that the ratio $f_\mathrm{corr}$ has a significantly smaller dynamic range than $w_{p,gg}^\mathrm{sim}$. Additionally, $w_{p,gg}^\mathrm{model}$ is capable of capturing a significant amount of the sensitivity to our HOD and cosmological parameters with insignificant computational expense. The upshot is that our emulation scheme is able to more accurately fit $f_\mathrm{corr}$ than $w_{p,gg}^\mathrm{sim}$. For galaxy-galaxy lensing, on the other hand, we have found that we can achieve acceptable modeling errors when emulating $\Delta \Sigma$ directly. This success may not hold for a different set of modeling requirements (e.g., a larger survey that yields smaller measurement errors), in which case we could emulate a similar halo-model correction for $\Delta \Sigma$.

To calculate $w_{p,gg}^\mathrm{model}$ we integrate over an analytically calculated real-space galaxy autocorrelation function $\xi_{gg}^\mathrm{model}$, which is expressed as a quadrature sum of 1- and 2-halo terms,
\beq
\xi_{gg}^\mathrm{model} (r) = \sqrt{  \left( \xi_{gg}^\mathrm{1h} \right)^2 + \left( \xi_{gg}^\mathrm{2h} \right)^2 }.
\eeq
The two-halo term is given by,
\beq
\xi_{gg}^\mathrm{2h} = b_g^2 \times \xi_{mm}(r),
\eeq
where $\xi_{mm}$ is the linear theory matter-matter correlation function and $b_g$ is the galaxy-bias calculated by integrating over the HOD and halo-mass function $d n_h / d M_h$ and halo bias function $b_h(M_h)$,
\beq
b_g = \frac{1}{n_\mathrm{gal}} \int_0^\infty d M_h \frac{d n_h}{d M_h} \left< N (M_h) \right> b_h(M_h).
\eeq
The more complicated 1-halo term is a sum of central-satellite $DD_{cs}$ and satellite-satellite $DD_{ss}$ pairs,
\beq
1 + \xi_{gg}(r) = \frac{DD_{cs}(r) + DD_{ss}(r)}{RR(r)},
\eeq
where $RR(r) = 2 \pi r^2 n_g$. These terms are written as,
\begin{align}
DD_{cs}(r) =& \int_0^\infty \langle N_\mathrm{cen} (M_h) \rangle \langle N_\mathrm{sat}(M_h) \rangle I^\prime \left( \frac{r}{r_h (M_h)} , c_\mathrm{vir}(M_h) \right)\\
&\times \frac{d n_h}{ d M_h} \frac{1}{r_h(M_h)} d M_h, \nonumber
\end{align}
\begin{align}
DD_{ss}(r) =& \frac{1}{2} \int_0^\infty \langle N_\mathrm{sat}(M_h) \rangle^2 F^\prime \left( \frac{r}{r_h (M_h)} , c_\mathrm{vir}(M_h) \right)\\
&\times \frac{d n_h}{ d M_h} \frac{1}{r_h(M_h)} d M_h, \nonumber
\end{align}
where $I^\prime$ and $F^\prime$ are dimensionless, differential pair count functions for an NFW profile. In the interest of brevity we omit expressions for these terms and direct the reader to the appendices of \citet{Wibking_et_al_2020} for them. In the calculation of the 1- and 2-halo terms we utilize the mass function $d n_h / d M_h$ of \citet{Tinker_et_al_2008}, the halo bias function $b_h(M_h)$ of \citet{Tinker_et_al_2010}, and the redshift dependent concentration-mass relation of \citet{Correa_2015}. The matter-matter correlation function $\xi_{mm}$ is obtained by Fourier transforming the linear matter power spectrum calculated with the fitting formula of \citet{Eisenstein_Hu_1998}. Although this calculation of $\xi_{gg}^\mathrm{model}$ would not be accurate enough on its own for DES analysis, it allows us to construct a high accuracy emulator.

\subsection{Emulation using Gaussian Processes}

\begin{figure*}
\centering
\includegraphics[width=1.0\textwidth]{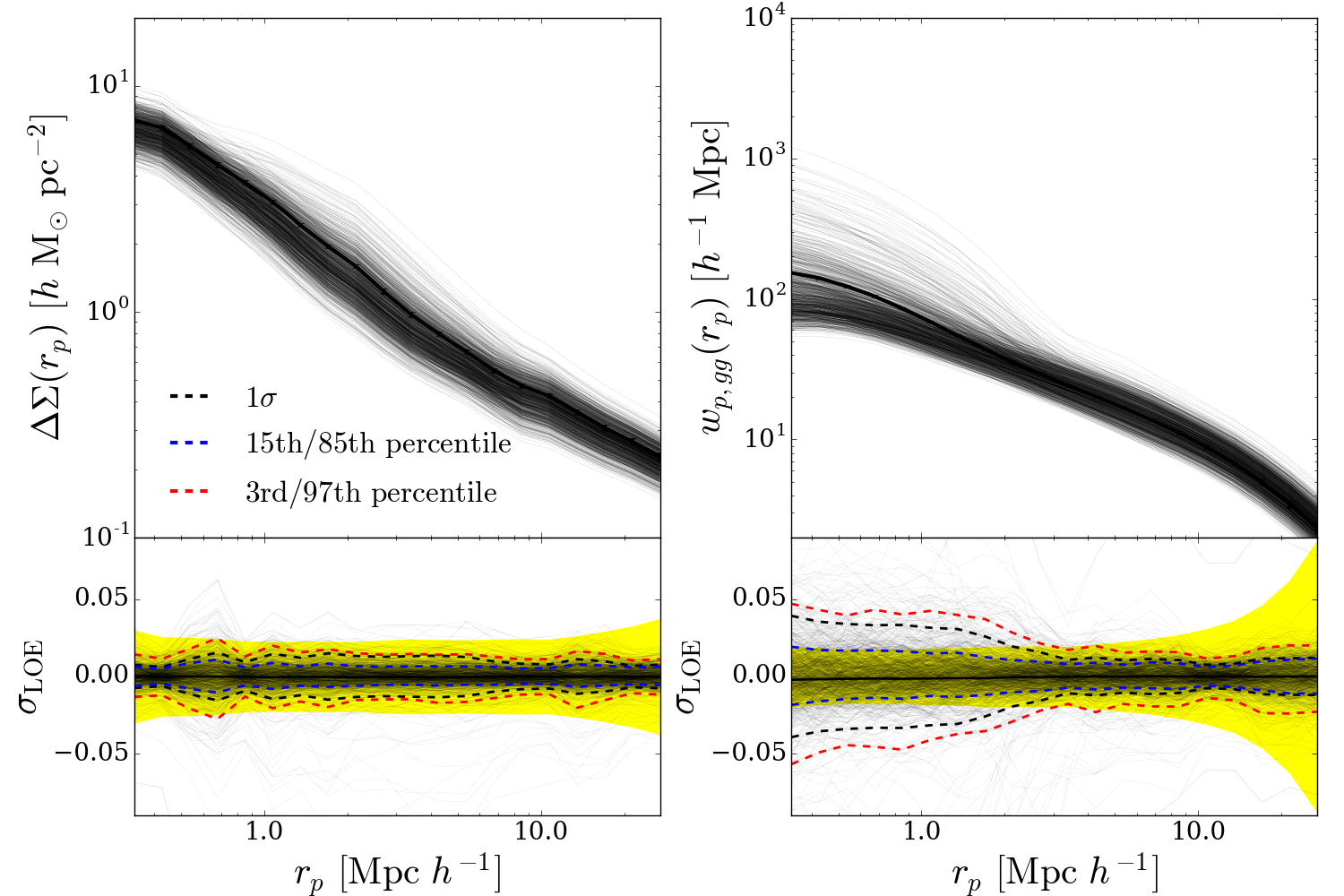}
\caption{$\Delta \Sigma$ (left) and $w_{p,gg}$ (right) as predicted by our simulations in our $z{=}0.5$ bin. Each of the grey-scale lines plots one of the 1000 training HOD+cosmology models we use to compute our emulator, while the black line shows the prediction for our fiducial model. The respective bottom panels show the leave-one-out emulator error for our two observables compared with the predicted observable covariance, plotted as a yellow band.}
\label{fig:emulator}
\end{figure*}

To model the dependence of $w_{p,gg}$ and $\Delta \Sigma$ on cosmological and HOD parameters, we implement the Gaussian process emulation scheme of \citet{Wibking_et_al_2020}. This amounts to performing a Gaussian process regression with a squared-exponential kernel in each radial bin of $f_\mathrm{corr}{=}w_{p,gg}^\mathrm{sim} / w_{p,gg}^\mathrm{model}$ and $\Delta \Sigma$. In each bin the hyperparameters of the kernel are obtained by maximizing the leave-one-out cross validation pseudo-likelihood. For more details on this process we direct the reader to the relevant appendices in \citet{Wibking_et_al_2020}.

The input data for this emulation are computed from 1000 cosmology and HOD models obtained from assigning 25 randomly generated HOD models to each of the 40 AbacusCosmos simulations described in section \ref{sec:sims}. The cosmological models used in these 40 simulations are drawn from constraints from CMB, BAO and SN data \citep{AndersonBOSS_et_al_2014} using the Latin hypercube sampling method of \citet{Heitmann_et_al_2009}. To obtain HOD parameterizations we Latin hypercube sample over flat probability density distributions in the ranges given in table \ref{table:params}. While not strictly necessary in a forecast context, these ranges are chosen to produce a large volume in parameter space to demonstrate the utility of our emulation scheme. Also listed in table \ref{table:params} are fiducial values of each parameter chosen to roughly describe the high density sample of redMaGiC galaxies. We use these fiducial values to compute derivatives with respect to our combined HOD and cosmological parameter vector.  We produce 1000 HOD parameterizations using this method and randomly assign 25 to each of the 40 AbacusCosmos cosmologies without replacement. For each combination of HOD and cosmology we compute $f_\mathrm{corr}$ and $\Delta \Sigma$ as described in section \ref{sec:obs}. The set of 1000 $f_\mathrm{corr}$ and $\Delta \Sigma$ models serves as the input to our Gaussian process emulator.

In figure \ref{fig:emulator} we show the input data-set for emulation (top panels) of $f_\mathrm{corr}$ and $\Delta \Sigma$ as well as the respective modeling errors we obtain (bottom panels). The left panels correspond to $\Delta \Sigma$ while the right panels correspond to $w_{p,gg}$. Each top panel shows the fiducial model in black with 1000 additional faint lines representing the models we use to construct our emulator. We can see that the parameter space we model covers a large range in amplitude for both of our observables. Each bottom panel shows the respective leave-one-out-simulation error. This error is computed by training the emulator with all elements of the training set except for those associated with one of the 40 AbacusCosmos boxes, then comparing it to the observable computed with that simulation (cosmology) and HOD parameterization. This functions as a conservative estimate of the accuracy of our emulator. The yellow region in each panel shows the diagonal errors we assume for our forecast of DES-Y6 clustering and galaxy-galaxy lensing. We can see that for both $w_{p,gg}$ and $\Delta \Sigma$ our modeling errors are comparable to the statistical errors. For $w_{p,gg}$ we can see that the $1\sigma$ errors are noticeably larger than the 15th/85th error percentiles at small scales indicating that the errors are non-Gaussian. The outliers in error space are also outliers in $w_{p,gg}$ space, so the 15th/18th percentile curves are more indicative of modeling errors that would appear in a likelihood analysis of data. Furthermore, the range of models being emulated is much larger than the $\pm5\sigma$ range expected from the DES errors, and training and applying the emulator over a more restricted range compatible with the measurements would yield still smaller emulator errors. We conclude that the current emulator is probably accurate enough to model $\Delta \Sigma$ and $w_{p,gg}$ in the final DES data, at least within our adopted parametric model, though further testing in the context of the final measurements will be desirable.

\subsection{Cosmological and HOD Derivatives}
\label{sec:deriv}

\begin{figure*}
\centering
\includegraphics[width=1.0\textwidth]{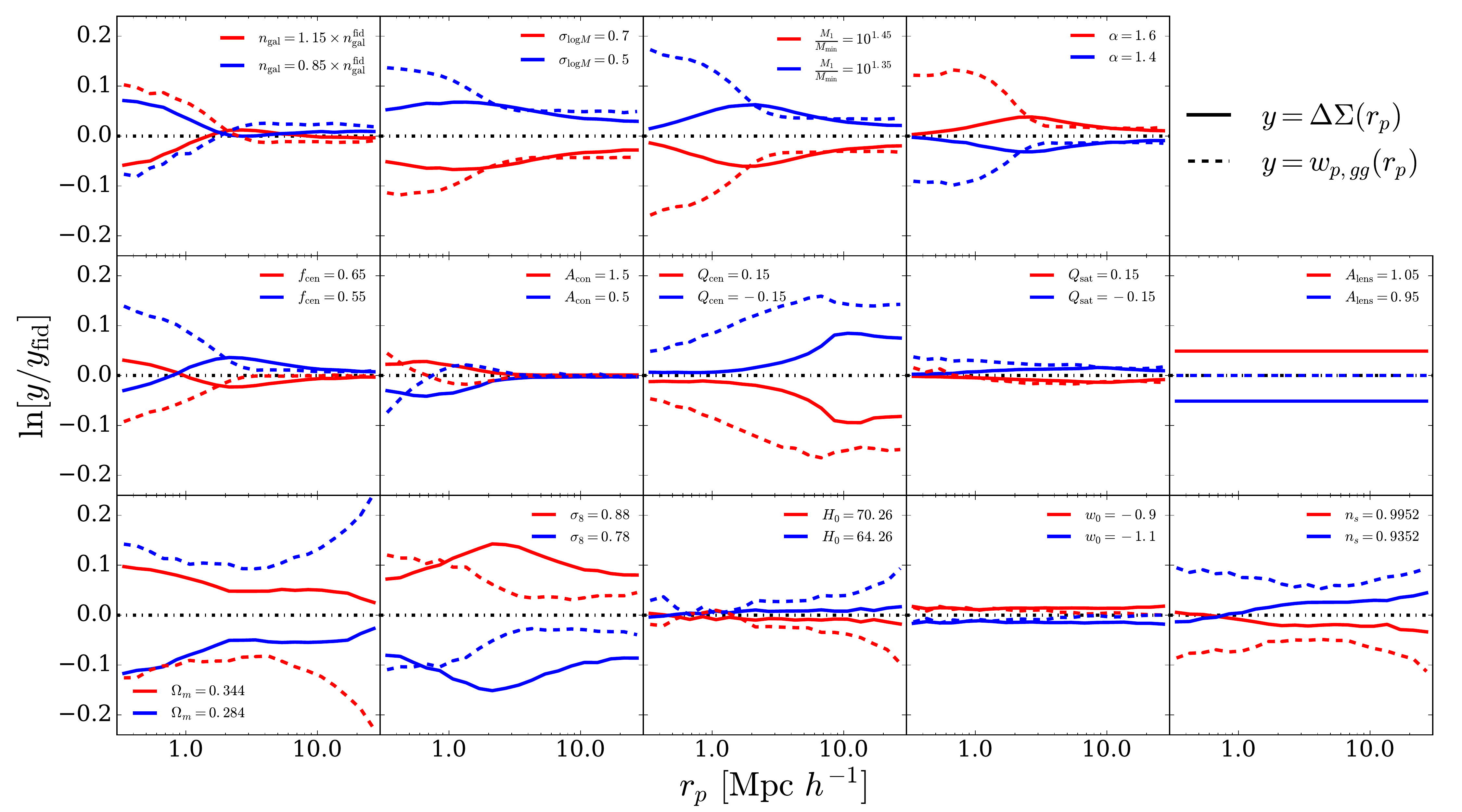}
\caption{Fractional changes to $\Delta \Sigma$ (solid lines) and $w_{p,gg}$ (dashed lines) induced by changes in HOD and cosmological parameters at $z=0.5$. In each panel red and blue curves show the emulator-predicted change of the observable for the parameter values indicted in the panel legend while holding all other parameters fixed.}
\label{fig:variations}
\end{figure*}

To compute derivatives for use in our Fisher forecast analysis, we use our emulator to compute $w_{p,gg}$ and $\Delta \Sigma$ at the fiducial values listed in table \ref{table:params} and steps up and down in each of our parameters. When using these derivatives to compute forecast constraints we additionally smooth them with a Savitsky-Golay filter. Rather than plot the derivatives directly, in figure \ref{fig:variations} we instead examine the impact of fixed variations in parameters for $w_{p,gg}$ and $\Delta \Sigma$. Curves in the figure are computed using simulations (snapshots at $z=0.5$). In each panel red (blue) curves show the effect of increasing (decreasing) the indicated parameter relative to the fiducial value for $\Delta \Sigma$ (solid lines) and $w_{p,gg}$ (dashed lines). 

We begin by examining the effects of parameter changes on $\Delta \Sigma$. We see that a decrease in the galaxy number density $n_\mathrm{gal}$ corresponds to an increase in $\Delta \Sigma$ at all scales. Recall that we treat $\ngal$ as an adjustable HOD parameter, not $M_\mathrm{min}$, so when we reduce $\ngal$ we increase $\Mmin$ and $M_1$ at fixed $M_1 / \Mmin$ to achieve the new density. At small scales this has a significant effect on $\Delta \Sigma$ by increasing the mean mass of halos that host galaxies. At large scales this increase in mean host halo mass leads to a small scale-independent increase in the galaxy bias. Turning to $\siglogM$ we see that an increase in the parameter decreases $\Delta \Sigma$ at all scales. At large scales this behavior is similar to the case of raising $n_\mathrm{gal}$: an increase in $\siglogM$ corresponds to a decrease in the mean host halo mass and therefore the galaxy bias. At small scales the increase in $\siglogM$ leads to a decrease in the satellite fraction. This small scale sensitivity exhibits an interesting scale dependence, peaking around $1.0{-}2.0\,\Mpch$, due to the offset in the positions of satellites and the peak of the matter distribution at the center of the host halo.

The parameters $\satfrac$ and $\alpha$ exhibit very similar behavior. Increasing (decreasing) $\alpha$ ($\satfrac$) increases the galaxy bias and satellite fraction, leading to a boost to the amplitude of $\Delta \Sigma$ at all scales. We note that the extent to which the satellite fraction is increased due to changes in $M_1$ and $M_\mathrm{min}$ at fixed $\satfrac$ depends on the value of $\alpha$. Because $\alpha$ is relatively high, with a fiducial value of $1.5$, a decrease in $M_1$ contributes relatively more satellites than an equal decrease in $M_\mathrm{min}$ contributes centrals. Depending on the shape of the halo mass function, a lower value of $\alpha$ could reverse this situation.

The remaining HOD parameters $\fcen$, $\Acon$, $\Qcen$, and $\Qsat$ exhibit more interesting behavior. When $\fcen$ is decreased this boosts the large scales of $\Delta \Sigma$ by increasing the satellite fraction (since $\ngal$ is held fixed) and therefore the galaxy bias. However at the smallest scales this leads to a decrease in $\Delta \Sigma$. This is because central galaxies residing in the most massive halos contribute significantly to $\Delta \Sigma$. Unlike changes in the satellite fraction due to $\siglogM$ or $\satfrac$, changing $\fcen$ removes some of these high-signal central galaxies. An increase in the parameter $\Acon$ increases the concentration of satellite galaxies. This moves satellites closer to the peak of the matter distribution within halos and therefore increases the 1-halo term of $\Delta \Sigma$. Because $\Acon$ does not affect the mean occupation at all, it has no effect on large scales. 

Turning to the first of our assembly bias parameters we see that a decrease in $\Qcen$ boosts $\Delta \Sigma$ at all scales. This is because negative values of $\Qcen$ decrease $M_\mathrm{min}$ for haloes in dense environments. This leads to a significant increase in $\Delta \Sigma$ at large scales peaking around $8.0 \, \Mpch$. The effect decreases towards small scales, though it does not vanish. Similarly a decrease in $\Qsat$ decreases $M_1$ for haloes in dense environments, boosting $\Delta \Sigma$ at all scales. Interestingly $\Qsat$ has a much smaller effect on $\Delta \Sigma$ than $\Qcen$. This is because the variation in bias for low mass haloes that may host a central is much larger than for the high mass haloes that host satellites.

Next are our cosmological parameters. We see that an increase in $\Omega_m$ leads to an increase in $\Delta \Sigma$ at all scales, with some mild scale dependence at small scales. The effect of increasing $\Omega_m$ on the linear power spectrum is to shift it towards higher $k$, or equivalently to shift $\xi_{mm}$ towards lower $r$.  This leads to a decrease in $\xi_{mm}$ at large scales and an increase at small scales that is suppressed by non-linear evolution.  The large scale decrease is counteracted and overcome by the increase in the $\Omega_m \rho_\mathrm{crit}$ prefactor in $\Delta \Sigma$, and the increase in $\Delta \Sigma$ at small scales is larger still. Increasing $\sigma_8$ also increases $\Delta \Sigma$ at all scales though with a different scale dependence that peaks at $r_p \approx 2.0 - 3.0 \, \Mpch$. This is directly due to an increase in $\xi_{mm}$ at all scales. An increase in $H_0$, like an increase in $\Omega_m$, shifts the linear power spectrum towards higher $k$, and the impact on the $\Omega_m \rho_\mathrm{crit}$ prefactor is absorbed by measuring $\Delta \Sigma$ in units of $h^{-1} M_\odot \mathrm{pc}^2$. The impact is a small decrease in $\Delta \Sigma$ at large $r_p$. Note that when we compute derivatives with respect to $\sigma_8$ we use the value of $\sigma_8$ at $z=0$ rather than the relevant snapshot redshift. In principle this choice can affect our constraints, but in practice the effect is small. 

Increasing $w_0$ from $-1.0$ to $-0.9$ leads to a slight scale independent increase in $\Delta \Sigma$, which is due to the analogous increase in $\xi_{mm}$. With $w_0 = -0.9$, structure growth ``freezes'' at slightly higher redshift, and with $\sigma_8$ fixed at $z=0$ the implied clustering at $z>0$ is larger.
Increasing $n_s$ makes the linear power spectrum bluer, decreasing the large scale $\xi_{mm}$ (and thus $\Delta \Sigma$) relative to the $8.0 \, \Mpch$ scale. The converse effect on small scales is damped by non-linear evolution.

Turning to $w_{p,gg}$ we see similar behavior as $\Delta \Sigma$ for many of our HOD parameters, particularly at large scales. As discussed in the $\Delta \Sigma$ case, the most important effect an HOD parameter can have on large scales is to change the galaxy bias by changing the mean host halo mass. We can see this in the case of $n_\mathrm{gal}$, $\siglogM$, $\satfrac$, $\alpha$, and $\fcen$, where the effects on the large scales of $w_{p,gg}$ of our parameter variations are qualitatively similar to those for $\Delta \Sigma$. There are subtle differences in scale dependence at large scales due to the fact that $\Delta \Sigma$ is an {\emph{excess}} surface density rather than a local overdensity. At small scales there are more significant differences between $w_{p,gg}$ and $\Delta \Sigma$ because of the strong impact of the satellite fraction on $\xi_{gg}$ in the 1-halo regime. For $\ngal$ the small scale effect is opposite in sign to the large scale effect because the increase of satellite galaxies dominates over the reduction in the galaxy bias. Our next three parameters, $\siglogM$, $\satfrac$, and $\alpha$, exhibit similar small scale behavior because they all also increase the satellite fraction at fixed $\ngal$. 

In contrast the central incompleteness parameter $\fcen$ exhibits significantly different behavior at small scales than $\Delta \Sigma$. Because $w_{p,gg}$ includes a satellite-satellite contribution to the 1-halo term, increasing the satellite fraction with a reduced $\fcen$ at fixed $\ngal$ leads to an increase in the 1-halo term at all scales. Turning to $\Acon$ we see further differences in small scale behavior. An increase in $\Acon$ boosts the very smallest scales of $w_{p,gg}$ but has a compensatory decrease at larger scales still within the 1-halo term. This is because the pairs gained at small scales by sharpening the galaxy profile concentration are lost at larger scales. 

Our assembly bias parameters also exhibit different behavior than in the case of $\Delta \Sigma$. A decrease of $\Qcen$ increases $w_{p,gg}$ at all scales but does so more significantly than for $\Delta \Sigma$. We also observe the same peak around $8.0 \, \Mpch$, but it is much smoother. The satellite assembly bias parameter $\Qsat$ exhibits similar behavior for $w_{p,gg}$ and $\Delta \Sigma$ at large scales, but it differs at small scales. Interestingly, both a decrease and increase of $\Qsat$ boost the small scales of $w_{p,gg}$. This is because we have chosen for our fiducial value $\Qsat = 0.0$, which minimizes the value of the second moment of the halo occupation $\langle N^2_\mathrm{gal}  (M_h) \rangle$.

The effect of our cosmological parameters on $w_{p,gg}$ is in most cases similar to the effect on $\Delta \Sigma$, albeit with different detailed scale dependence. The notable exceptions are $\Omega_m$ and $n_s$ at small scales. Unlike the case of $\Delta \Sigma$, an increase in $\Omega_m$ leads to a decrease in $w_{p,gg}$ because there is no longer a prefactor proportional to $\Omega_m$. The small scale behavior follows from the upward shift in the halo mass function caused by higher $\Omega_m$. To achieve the same number density at fixed $\satfrac$, both $M_1$ and $M_\mathrm{min}$ must shift to high values. As discussed previously this leads to a decrease in the satellite fraction and therefore depresses the 1-halo term. The origin of the small scale impact of $n_s$ is not obvious, but we suspect it derives from the effect of the power spectrum shape on the halo mass function.

\subsection{Summary}

Regardless of the detailed explanations of each curve in figure \ref{fig:variations}, our critical finding is that each parameter that has a significant impact on $\Delta \Sigma$ or $w_{p,gg}$ does so with a distinct scale dependence, which is typically different for the two observables. Therefore, even though the HOD introduces many free parameters, precise measurements of $\Delta \Sigma$ and $w_{p,gg}$ over a wide dynamic range provide enough information to break parameter degeneracies and achieve tight constraints on cosmological parameters. The distinctive scale dependence arises because we span the linear, trans-linear, and fully non-linear regimes. Modeling measurements into small scales thus offers the prospect of significantly improving cosmological inferences from weak lensing and galaxy clustering data, as we demonstrate in subsequent sections.

\section{Covariance Estimation}
\label{sec:cov}

We use a combination of analytic and numerical methods to compute the observable covariance matrix for $w_{p,gg}$ and $\Delta \Sigma$. We analytically compute the $\Delta \Sigma$ covariance using a Gaussian formalism, i.e. assuming the galaxy and matter fields are Gaussian random and adding a shape noise contribution \citep[e.g.][]{Singh_et_al_2017, Wibking_et_al_2020}. Recently \cite{Wu_et_al_2019}, in the context of cluster weak lensing, showed that the standard Gaussian formalism for computing the lensing covariance becomes insufficient when the large-scale structure contribution to the covariance becomes comparable to shape noise. Because our $\Delta \Sigma$ covariance is shape-noise dominated we utilize the standard Gaussian formalism, but we note that in a deeper weak lensing survey than DES it may become insufficient for galaxy-galaxy lensing as well. Because the lensing covariance matrix is shape-noise dominated, we also ignore the cross-observable covariance with $w_{p,gg}$ and treat the two observables as independent in all that follows.

We include a correction to the $\Delta \Sigma$ covariance matrix to analytically marginalize over potential contributions from a point mass at the center of each galaxy lens \citep[e.g.][]{MacCrann_et_al_2020, Wibking_et_al_2020}. This enclosed point mass, which is allowed to be positive or negative, can represent the impact of small scale substructure that is unresolved and absent from our simulations. It can also characterize the impact of baryonic physics effects like dissipation and feedback. In the covariance matrix, the point-mass correction takes the form
\beq
\tilde{C} = C + \sigma^2 v v^T,
\eeq
where $v$ is a column vector with values $\left[r_{p,0}^{-2}, r_{p,1}^{-2}, ... , r_{p,N}^{-2} \right]$ and $\sigma$ is the width of the Gaussian prior on the enclosed point mass.  We use the Sherman-Morrison matrix identity and assume a flat prior on $\sigma$ \citep[e.g.][]{MacCrann_et_al_2020, Wibking_et_al_2020}, yielding
\beq
\tilde{C}^{-1} = C^{-1} - \frac{C^{-1} v v^T C^{-1}}{v^T C^{-1} v}.
\eeq

To compute the covariance for $w_{p,gg}$ we use a combination of analytic and numerical methods. Unlike $\Delta \Sigma$ the covariance for $w_{p,gg}$ contains a significant non-Gaussian contribution, particularly at small scales. To account for this contribution we use bootstrap methods to numerically compute the covariance using the 20 $(1100.0 \; \Mpch)^3$ simulation boxes of a fiducial cosmology with different phases from \citet{Garrison_et_al_2018}. Each box is divided into 25 equal area subvolumes in the $x-y$ plane. In each subvolume $w_{p,gg}$ is computed in projection for the fiducial HOD model. We obtain 500 bootstrap resamples by choosing 500 subvolumes with replacement and averaging $w_{p,gg}$ for each resample. These bootstrap resamples are used to compute the covariance for $w_{p,gg}$. This numerical covariance matrix is inherently noisy and may lead to optimistically biased forecasted parameter constraints. For this reason we also compute the Gaussian covariance for $w_{p,gg}$ \citep[e.g.][]{Cooray_Hu_2001, Marian_Smith_Angulo_2015, Krause_Eifler_2017, Singh_et_al_2017} and use the diagonal elements of the numerical covariance matrix to normalize the analytic correlation matrix. Thus, our final covariance matrix uses the Gaussian model to compute off-diagonal correlations and the numerical simulations to compute variances and to scale correlations to covariances.

Our forecasts are meant to model DES weak lensing and galaxy clustering with redMaGiC selected galaxies. Consequently we consider three bins of redshift for our galaxies, $z = 0.15- 0.35$, $z = 0.35-0.55$ and $z = 0.55-0.75$, and we assume a survey area of $\Omega = 5000 \; \mathrm{deg}^2$. These bins are modeled using AbacusCosmos simulation snapshots at $z = 0.3$, $z = 0.5$, and $z=0.7$ respectively which are also assumed as lens redshifts when calculating $\Sigcrit$. Mean source redshifts are computed using the source redshift distribution of \citet{Rozo_2011}. This source redshift distribution is also used to compute source surface densities in each bin assuming a total source surface density of $\Sigma_\mathrm{src} = 10.0 \; \mathrm{arcmin^{-2}}$. We assume a shape noise per galaxy of $\sigma_\gamma = 0.2$.

\section{Cosmological Forecasts}
\label{sec:forecasts}

\subsection{Fiducial scenario}
\label{subsec:fid}

\begin{figure*}
\centering
\includegraphics[width=1.0\textwidth]{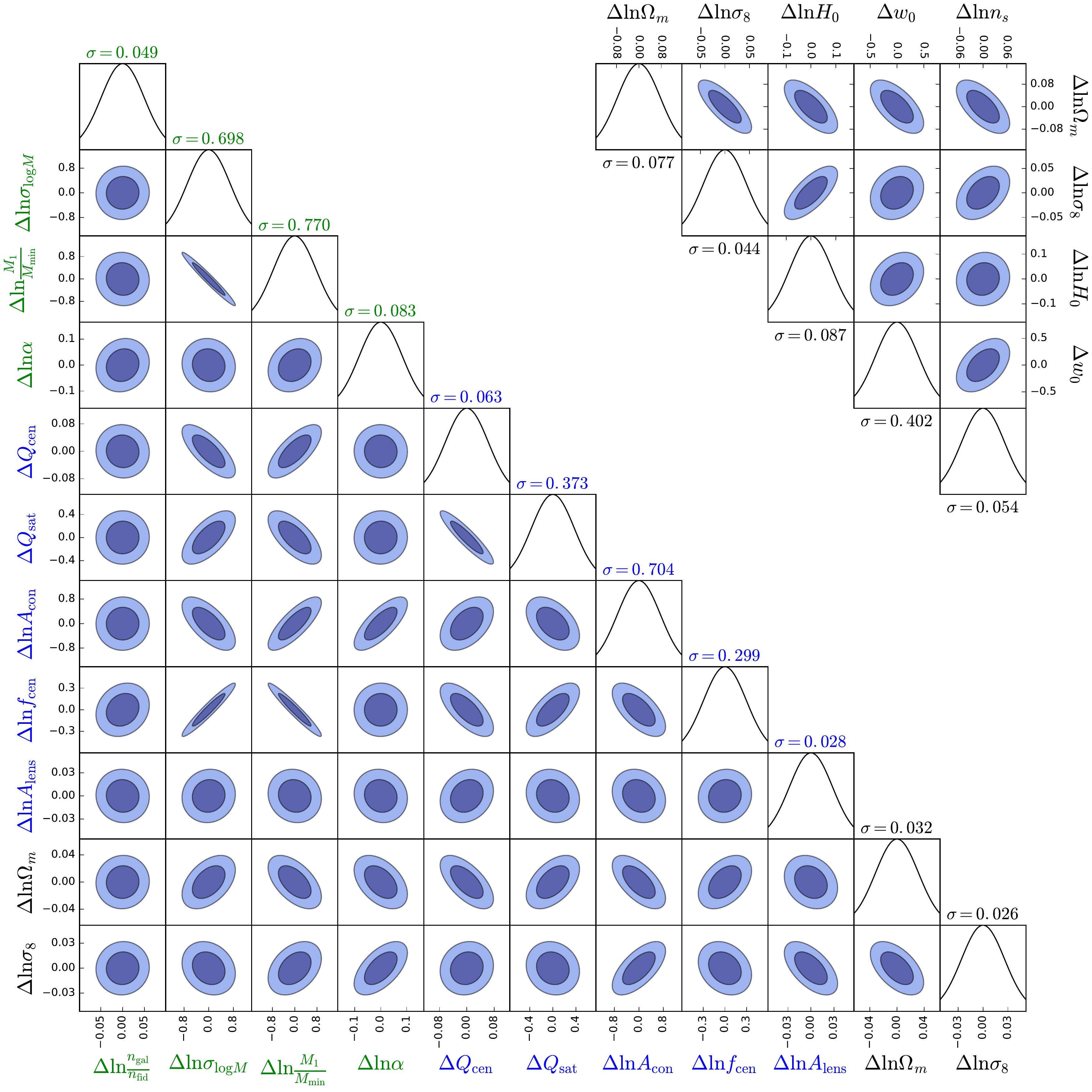}
\caption{Forecast parameter constraints ($68\%$ and $95\%$ contours) for our fiducial scenario, assuming DES-Y6 survey parameters for galaxies between $z{=}0.35-0.55$, and using all scales $0.3<r_p<30.0\,\Mpch$ of $\Delta \Sigma$ and $w_{p,gg}$. The bottom block shows constraints on $\Omega_m$ and $\sigma_8$ and all of our HOD parameters while holding all other cosmological parameters ($H_0$,$w_0$,$n_s$) fixed at their fiducial values. The upper right block shows constraints on all cosmological parameters while marginalizing over all HOD parameters. Fully marginalized errors on each parameter are listed above each PDF panel.}
\label{fig:corner}
\end{figure*}

We forecast parameter constraints for our fiducial scenario, a DES-like survey, with the covariance matrix described in section \ref{sec:cov} and derivatives calculated by finite difference from emulator predictions described in section \ref{sec:deriv}. Additionally we impose a $5\%$ Gaussian prior on the galaxy number density and a $3\%$ prior on $\Alens$. The parameter $\Alens$  allows for some amount of scale independent lensing bias. It can be thought of as representing some combination of uncertainty in shear calibration and photometric redshift errors that lead to uncertainty in $\Sigcrit$. Our choice of a $3\%$ prior on $\Alens$ is loosely motivated by \citet{MacCrann_DES_et_al_2020} and \citet{Myles_DES_et_al_2021}. Note that we forecast constraints for the natural logarithm of our parameters, except for $\Qcen$ and $\Qsat$ which can be zero or negative. 

Our fiducial scenario combines $w_{p,gg}$ and $\Delta \Sigma$ with information from $0.3 \; \Mpch < r_p < 30.0 \Mpch$ in the $z = 0.35 - 0.55$ redshift bin. We focus on this single redshift bin for the sake of clarity, chosen because it produces the strongest constraints. In section \ref{subsec:redshift} we examine constraints from our other redshift bins. Results for this fiducial case are shown in figure \ref{fig:corner}. The bottom left block shows our forecast with all cosmological parameters other than $\Omega_m$ and $\sigma_8$ fixed. The upper right block shows the fiducial constraints on all cosmological parameters (note that these constraints are marginalized over all other HOD and nuisance parameters, which have been suppressed for visual clarity). Typically other data, such as CMB anisotropies, the supernova Hubble diagram, and the galaxy power spectrum, provide tight constraints on $H_0$, $w_0$, and $n_s$, so the fixed parameter case is more representative of what DES can achieve on $(\Omega_m, \sigma_8)$ in a multi-probe analysis.

When $\sigma_8$ and $\Omega_m$ are our only cosmological parameters the best constrained combination of the two is $\sigma_8 \Omega_m^{0.438}$, with a $1\sigma$ uncertainty of $2.19\%$ after marginalizing over the halo-galaxy connection. Individual marginalized constraints on $\sigma_8$ and $\Omega_m$ are $2.6\%$ and $3.2\%$. Our choice to constraint $\sigma_8(z{=}0)$ rather than $\sigma_8(z{=}0.5)$ affects the $\Omega_m{-}\sigma_8$ constraint slightly because of the effect of $\Omega_m$ on the growth factor, but the effect is smaller than our precision. For example, a $3.0\%$ difference in the value of $\Omega_m$ corresponds to a sub-percent change in the linear growth factor at $z{=}0.5$. There are significant degeneracies between $\sigma_8$ and HOD or nuisance parameters, particularly $\Alens$ and $\alpha$. In the case of $\Omega_m$ there is a significant degeneracy with $\Qcen$, likely due to the large scales of $w_{p,gg}$. Among HOD parameters $\siglogM$ and $\satfrac$ exhibit a strong degeneracy, leading to poor constraints on both parameters. This is unsurprising as both parameters have virtually the same effect on both of our observables (see figure \ref{fig:variations}). Interestingly $\fcen$ also exhibits a strong degeneracy with both $\siglogM$ and $\satfrac$, likely due to the way all three parameters affect $w_{p,gg}$. Our two assembly bias parameters, $\Qcen$ and $\Qsat$ also exhibit a strong degeneracy with each other, likely due to their similar scale dependence at large $r_p$. Constraints on $\Qcen$ are much tighter than constraints on $\Qsat$ because it has a much stronger effect on our observables.

When we forecast with all other cosmological parameters free we find constraints of $4.4\%$ and $7.7\%$ on $\sigma_8$ and $\Omega_m$, a degradation by roughly a factor of 2. In this case the best constrained combination of $\sigma_8$ and $\Omega_m$ is $\sigma_8 \Omega_m^{0.444}$ with a forecasted constraint of $2.79\%$, moderately degraded from the $2.19\%$ constraint with fixed $H_0$, $w_0$, and $n_s$. With DES data alone, much of the ability to break the $\Omega_m{-}\sigma_8$ degeneracy comes from the shape of $w_{p,gg}$, but the impact of $H_0$ and $n_s$ on the linear power spectrum is largely degenerate with that of $\Omega_m$. Leaving these parameters free therefore widens the constraints on $\Omega_m$ and $\sigma_8$ individually but with less impact on their best constrained combination. The value of $w_0$ has little impact on our observables (figure \ref{fig:variations}), and unsurprisingly we do not forecast a meaningful $w_0$ constraint. The value of $w_0$ is somewhat degenerate with $\sigma_8$ and $\Omega_m$ because $\sigma_8$ is defined at $z = 0$ and our observation redshift is $z=0.5$. If we fix $w_0$ but leave $H_0$ and $n_s$ free then the constraint on the best constrained parameter $\sigma_8$ and $\Omega_m$ combination $\sigma_8 \Omega_m^{0.604}$ improves to $2.3\%$, similar to the case with all three parameters fixed. In contrast the constraint on $\Omega_m$ only improves to $6.0\%$ compared to the $3.2\%$ constraint when $H_0$, $n_s$, and $w_0$ are fixed. We discuss constraints in the $S_8 - \Omega_m$ plane below, for the fiducial scenario and other cases. 

\subsection{Impact of systematics: $\Alens$ and point-mass}
\label{subsec:systematics}

\begin{table*}
   \centering
   \caption{Parameter forecast uncertainties with $H_0$, $w_0$, and $n_s$ and fixed, in the $z=0.5$ bin.}
    \begin{tabular}{lcccc}
      \hline
        Case & $\Delta \ln \Alens$ & $\Delta \ln \Omega_m$ & $\Delta \ln \sigma_8$ & $\Delta \ln S_8$ \\
 	    \hline
 	    $\Delta \Sigma$ and $w_{p,gg}$, $\Alens$ fixed & - & 0.031 & 0.021 & 0.012\\
        $\Delta \Sigma$ and $w_{p,gg}$, $\Alens$ free, $3\%$ prior & 0.028 & 0.032 & 0.026 & 0.022\\
        $\Delta \Sigma$ and $w_{p,gg}$, $\Alens$ free, no prior & 0.078 & 0.037 & 0.047 & 0.053\\
        $\Delta \Sigma$ and $w_{p,gg}$, No point-mass, $\Alens$ fixed & - & 0.031 & 0.021 & 0.012\\
        $\Delta \Sigma$ and $w_{p,gg}$, No point-mass, $\Alens$ free, $3\%$ prior & 0.028 & 0.031 & 0.026 & 0.022\\
        $\Delta \Sigma$ and $w_{p,gg}$, No point-mass, $\Alens$ free, no prior & 0.078 & 0.036 & 0.047 & 0.053\\
        \hline
   \end{tabular}
   \label{table:margAlens-tests}
\end{table*}

Our forecasts include two important sources of systematic uncertainty in $\Delta \Sigma$. As described in section \ref{sec:cov} we modify our lensing covariance to marginalize over an enclosed point-mass. This point-mass marginalization is meant to characterize the impact of baryonic physics on the mass profile within halos as well as representing small scale substructure potentially unresolved by our simulations. We also include a multiplicative bias parameter $\Alens$ that captures potential scale-independent errors in lensing calibration. This may be caused by errors in shear calibration or errors in the measurement of $\Sigcrit$.

To test the sensitivity of constraints to these systematics we perform a variety of tests and list resulting forecasted constraints on $\Alens$, $\Omega_m$, $\sigma_8$ and $S_8 = \sigma_8 \Omega_m^{0.5}$ in table \ref{table:margAlens-tests}. In these tests we fix all cosmological parameters besides $\sigma_8$ and $\Omega_m$ and we marginalize over all HOD parameters. All of these results are for the $z=0.35-0.55$ bin only. Our first series of tests utilizes the full datavector ($\Delta \Sigma$ and $w_{p,gg}$). We see that our constraints on $\Alens$ are largely prior dominated; when $\Alens$ is free with no prior our datavector only constrains it at the $7.8\%$ level. This significantly degrades our forecast constraint on $\sigma_8$, almost doubling the uncertainty from $2.6\%$ to $4.7\%$, but it has less of an effect on $\Omega_m$.

Because the impact of parameters other than $\Alens$ is scale-dependent, we might hope that modeling $\Delta \Sigma$ and $w_{p,gg}$ into non-linear scales could break the degeneracy between $\Alens$ and cosmology. Table \ref{table:margAlens-tests} shows that this is only partly the case. If we adopt no prior on $\Alens$ then our data vector constrains it to $7.8\%$ and constrains $S_8$ to $5.3\%$. This is a huge improvement on linear theory, where $\Alens$ and $\sigma_8$ are perfectly degenerate. However, with a $3\%$ $\Alens$ prior, the posterior uncertainty in $\Alens$ is only slightly better at $2.8\%$. Furthermore, the $\Alens$ uncertainty remains a significant limitation, causing the $S_8$ uncertainty to be $2.2\%$ instead of the much stronger $1.2\%$ that could be achieved if $\Alens$ were known perfectly. We further examine the sensitivity of our constraints to our $\Alens$ prior in figure \ref{fig:Alens}, discussed below

We next test the robustness of our forecasts to our point-mass marginalization scheme. We repeat each of the previous tests without including this modification to the lensing covariance. We see that in this case the point-mass marginalization has very little effect on the final constraints. When $\Alens$ is fixed it has a completely negligible effect. When we assume a $3\%$ prior on $\Alens$ or assume no prior, the point-mass marginalization has a very small effect on constraints on $\Omega_m$. These results suggest, that for our data-vector, the small scales of $\Delta \Sigma$ are not the most important regime for constraining $\Omega_m$ or $\sigma_8$. It may also appear to suggest that the point-mass marginalization is unimportant, but we caution that this depends on the choice of data-vector and galaxy sample. Because our assumed lensing covariance is shape-noise dominated, we can imagine a future scenario in which the errors on $\Delta \Sigma$ are substantially improved relative to $w_{p,gg}$. Conversely, a sparser lens sample would have larger errors for both $\Delta \Sigma$ and $w_{p,gg}$, but the impact on $w_{p,gg}$ could be larger. In either scenario, including marginalization over a point-mass would be more important because of the increased relative importance of the small scales of $\Delta \Sigma$. Also, while we are considering point mass marginalization as a proxy for baryonic physics uncertainties, it is necessary to check that it does in fact remove biases from baryonic effects at the DES statistical precision.

\subsection{Relative contributions of scales}
\label{subsec:scales}

\begin{table*}
   \centering
   \caption{Parameter forecast uncertainties with $H_0$, $w_0$, and $n_s$ fixed in the $z = 0.5$ redshift bin. Also included are constraints on $S_8 = \sigma_8 \Omega_m^{0.5}$. Entries in the first and second columns indicate which scales are {\emph{retained}} in the $\Delta \Sigma$ and $w_{p,gg}$ datavectors, with ``small'' indicating $0.3 - 3.0 \, \Mpch$ and ``large'' indicating $3.0 - 30.0 \, \Mpch$. All cases assume a $3\%$ prior on $\Alens$ and marginalization over a point mass contribution to $\Delta \Sigma$.}
   \resizebox{\textwidth}{!}{\begin{tabular}{llccccccccccccc}
      \hline
        $\Delta \Sigma $ & $w_{p,gg}$ & $\Delta \ln \frac{\ngal}{\nfid}$ & $\Delta \ln \sigma_{\log M}$ & $\Delta \ln \frac{M_1}{M_{\mathrm{min}}}$ & $\Delta \ln \alpha$ &$\Delta \Qcen$ & $\Delta \Qsat$ & $\Delta \ln \Acon$ & $\Delta \ln \fcen$ & $\Delta \ln \Alens$ & $\Delta \ln \Omega_m$ & $\Delta \ln \sigma_8$ & $\Delta \ln S_8$\\
 	    \hline
        all & all & 0.049 & 0.698 & 0.334 & 0.083 & 0.063 & 0.373 & 0.704 & 0.299 & 0.028 & 0.032 & 0.026 & 0.022\\
        \hline
        all & - & 0.050 & 3.353 & 0.990 & 0.709 & 0.275 & 3.234 & 5.763 & 1.095 & 0.030 & 0.354 & 0.233 & 0.238\\
        - & all & 0.050 & 1.649 & 0.678 & 0.162 & 0.123 & 0.613 & 1.108 & 0.564 & 0.030 & 0.061 & 0.184 & 0.206\\
        \hline
        small & all & 0.050 & 1.185 & 0.521 & 0.095 & 0.105 & 0.472 & 0.803 & 0.468 & 0.030 & 0.037 & 0.028 & 0.027\\
        large & all & 0.050 & 0.858 & 0.394 & 0.138 & 0.066 & 0.422 & 1.032 & 0.350 & 0.029 & 0.036 & 0.034 & 0.027\\
        \hline
        all & small & 0.050 & 0.825 & 0.431 & 0.139 & 0.110 & 0.899 & 1.172 & 0.359 & 0.029 & 0.065 & 0.053 & 0.029\\
        all & large & 0.050 & 1.439 & 0.594 & 0.190 & 0.085 & 0.522 & 0.846 & 0.554 & 0.029 & 0.036 & 0.037 & 0.030\\
        \hline
        small & small & 0.050 & 1.558 & 0.734 & 0.171 & 0.250 & 1.815 & 1.435 & 0.638 & 0.030 & 0.125 & 0.079 &  0.042\\
        large & large & 0.050 & 2.499 & 1.078 & 0.628 & 0.130 & 0.636 & 4.756 & 0.843 & 0.029 & 0.040 & 0.056 &  0.046\\
        \hline
   \end{tabular}}
   \label{table:uncertainties}
\end{table*}

\begin{figure}
\centering
\includegraphics[width=0.45\textwidth]{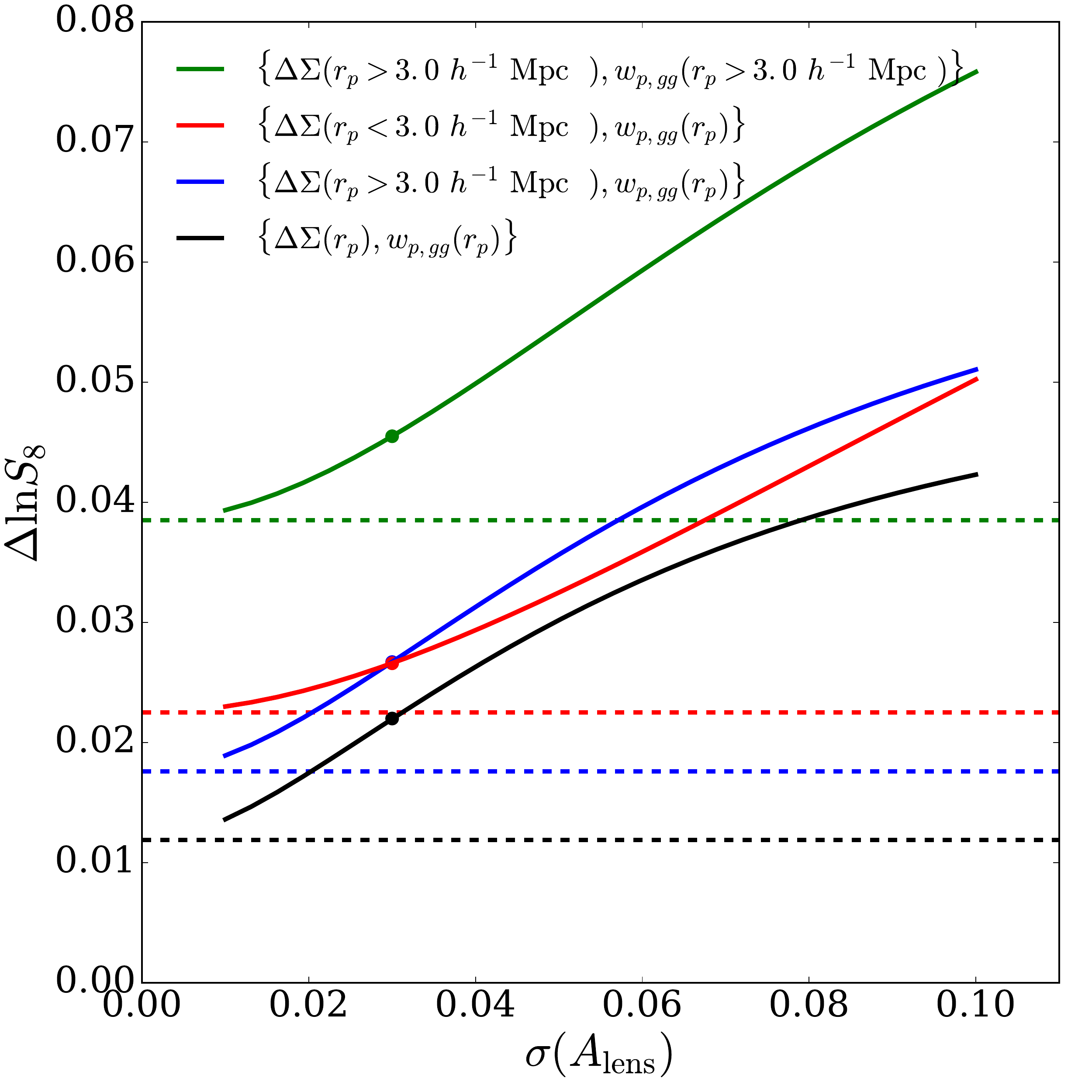}
\caption{Forecast constraints on $\ln S_8$ as a function of $\Alens$ prior in the $z{=}0.5$ bin marginalized over all HOD parameters with all other cosmological parameters fixed. The black line shows results from all scales of $\Delta \Sigma$ and $w_{p,gg}$, the red (blue) line shows results from small (large) scales of $\Delta \Sigma$ with all scales of $w_{p,gg}$, and the green line shows the results from large scale of both $\Delta \Sigma$ and $w_{p,gg}$. Points on these lines mark our fiducial forecast assumption of $\sigma(\Alens) = 0.03$. Analogously coloured dashed lines show constraints with $\Alens$ fixed.}
\label{fig:Alens}
\end{figure}

Table \ref{table:uncertainties} examines a variety of alternative scenarios in which we omit different elements of the fiducial datavector. In all of these tests we fix all cosmological parameters other than $\Omega_m$ and $\sigma_8$ and report constraints on $S_8 = \sigma_8 \Omega_m^{0.5}$. We assume our fiducial $3\%$ prior on $\Alens$ and $5\%$ prior on $\ngal$, and we include the point mass marginalization term in the lensing covariance matrix. In addition to omitting one of $\Delta \Sigma$ and $w_{p,gg}$ entirely, we also try omitting small ($r_p < 3.0 \; \Mpch$) and large ($r_p > 3.0 \; \Mpch$) scales of either. The choice of $3.0 \; \Mpch$ roughly corresponds to a division between the linear regime and non-linear regime, and it also splits each observable into equal numbers of data points. The first line of table \ref{table:uncertainties} (`all all') corresponds exactly to the fiducial scenario shown in figure \ref{fig:corner}. We again focus on the $z=0.5$ redshift bin for simplicity.

The second line of table \ref{table:uncertainties} shows a forecast with $\Delta \Sigma$ as the only observable. We see that the precision on all parameters has degraded drastically, except for $\ngal$ and $\Alens$ which have informative priors. Compared to the fiducial case, the precision on $\sigma_8$, $\Omega_m$, $S_8$ degrades by roughly a factor of ten. This poor performance is unsurprising: without galaxy clustering, galaxy-galaxy lensing in the linear regime has no cosmological constraining power because of degeneracy between $b_g$ and $\sigma_8$, and non-linear scale-dependence at DES measurement precision allows only moderate degeneracy breaking. 

We next consider the case of $w_{p,gg}$ on its own. We again see that all parameter constraints are significantly degraded, although not by as much as in the case of $\Delta \Sigma$ on its own. Constraints on all of our HOD parameters are significantly worse than with the full datavector, but are significantly better than from $\Delta \Sigma$ on its own. This difference is not surprising in the context of figure \ref{fig:variations}, which shows that $w_{p,gg}$ is generally more sensitive to the galaxy-halo connection, particularly at small scales. Since many of these HOD parameters are degenerate with each other, these individual improvements in sensitivity synergize with each other to significantly improve overall constraints on the HOD. The large scale shape of $w_{p,gg}$ constrains $\Omega_m$, so the cosmological parameter constraints from $w_{p,gg}$ alone are better than those from $\Delta \Sigma$ alone. However, fractional errors in $\sigma_8$ and $S_8$ are still at the $20\%$ level, drastically worse than the fiducial scenario. In linear theory the impact of $b_g$ and $\sigma_8$ on $w_{p,gg}$ would be fully degenerate. Non-linear scaling provides enough leverage to obtain $20\%$ precision, but $\sigma_8$ remains significantly degenerate with HOD parameters. As expected, precise constraints on matter clustering require both $\Delta \Sigma$ and $w_{p,gg}$.

The remaining lines of table \ref{table:uncertainties} show the impact of omitting small or large scale measurements from one or both components of the datavector. When we omit the large scales of $\Delta \Sigma$ (line 4, `small all'), the $S_8$ constraint degrades to $2.7\%$ from its fiducial value of $2.2\%$. Both $\sigma_8$ and $\Omega_m$ are individually degraded. If we retain the large scales of $\Delta \Sigma$ {\emph{instead}} of the small scales (line 5, `large all') then the $S_8$ precision is again $2.7\%$. The fact that large and small scales of $\Delta \Sigma$ can independently give precise $S_8$ constraints in concert with $w_{p,gg}$ has encouraging implications. Modeling systematics and some measurement systematics are likely to be very different in these two regimes, so comparing inferred parameters will provide a strong test of robustness and a valuable diagnostic of systematics if they are present.

If we retain all scales of $\Delta \Sigma$ but use only the small or large scales of $w_{p,gg}$ then $S_8$ constraints degrade to $2.9\%$ or $3.0\%$, respectively. HOD constraints are typically much worse if we have only large scales of $w_{p,gg}$, so it may seem surprising that $S_8$ constraints are comparable. However, in the linear regime it is only the overall galaxy bias factor $b_g$ that matters, so large trade-offs among HOD parameters may not have much impact on $S_8$ precision. Furthermore, the large scales of $w_{p,gg}$ provide better $\Omega_m$ constraints, so the breaking of $\Omega_m{-}\sigma_8$ degeneracy is considerably better for the `all large' scenario than the `all small' scenario.

The final rows of table \ref{table:uncertainties} show cases in which we take either large or small scales of both $w_{p,gg}$ and $\Delta \Sigma$. The most important takeaway is the large gain in cosmological constraining power from using all scales of $\Delta \Sigma$ and $w_{p,gg}$ (first line of table \ref{table:uncertainties}) versus using only scales $r_p > 3.0 \, \Mpch$ (last line). The improvement on $S_8$ precision from $4.6\%$ to $2.2\%$ is equivalent to a $(4.6/2.2)^2 \approx 4.4$ increase in survey area. The `small small' scenario slightly outperforms the `large large' scenario, with a $4.2\%$ versus $4.6\%$ $S_8$ precision. However, given the increased modeling complexity of on small scales there is no reason to contemplate pursuing this scenario in practice, whereas the `large large' scenario (in multiple redshift bins) is roughly analogous to the DES key project analyses performed to date.

We summarize and expand upon some of these results in figure \ref{fig:Alens}. Curves show the constraint on $S_8$ marginalized over the HOD with all other cosmological parameters fixed as a function of the prior assumed on $\Alens$. Each colour corresponds to a different forecast scenario from table \ref{table:uncertainties}, and analogous dashed lines show the constraint on $S_8$ when $\Alens$ is fixed. The black curve in figure \ref{fig:Alens} shows the relationship between the $\Alens$ prior and $S_8$ when the full data-vector is used. We see that if $\Alens$ were perfectly known then the best constraint we could achieve with our $z{=}0.5$ datavector is about a factor of two narrower than our fiducial scenario, $1.2\%$ versus $2.2\%$. At large values of $\sigma(\Alens)$ the curve begins to flatten around $\sigma(\Alens) = 0.07$ and asymptote towards a ${\sim}5\%$ constraint on $S_8$. This behavior is consistent with our results in table \ref{table:margAlens-tests}; when $\Alens$ is completely free our full datavector yields a $7.8\%$ constraint on $\Alens$ and a $5.3\%$ constraint on $S_8$. The red and blue curves correspond to omitting the large and small scales of $\Delta \Sigma$ respectively. Finally the green curve shows results when we omit the small scales of both of our observables. The relative ordering of these curves at a given $\sigma(\Alens)$ indicates the relative importance of the respective elements of the datavector. Given our fiducial prior on $\Alens$, the large and small scales of $\Delta \Sigma$ have similar impact on the constraint on $S_8$. The large difference between the black and green curves emphasizes the value of the small scales of both observables. If the $\sigma(\Alens)$ prior could be tightened from $0.03$ to $0.01$ then the difference between the all scales and large scales analysis would be equivalent a to $(3.93/1.36)^2 \approx 8.35$ times increase in survey area.

\subsection{Dependence on redshift}
\label{subsec:redshift}

\begin{figure}
\centering
\includegraphics[width=0.45\textwidth]{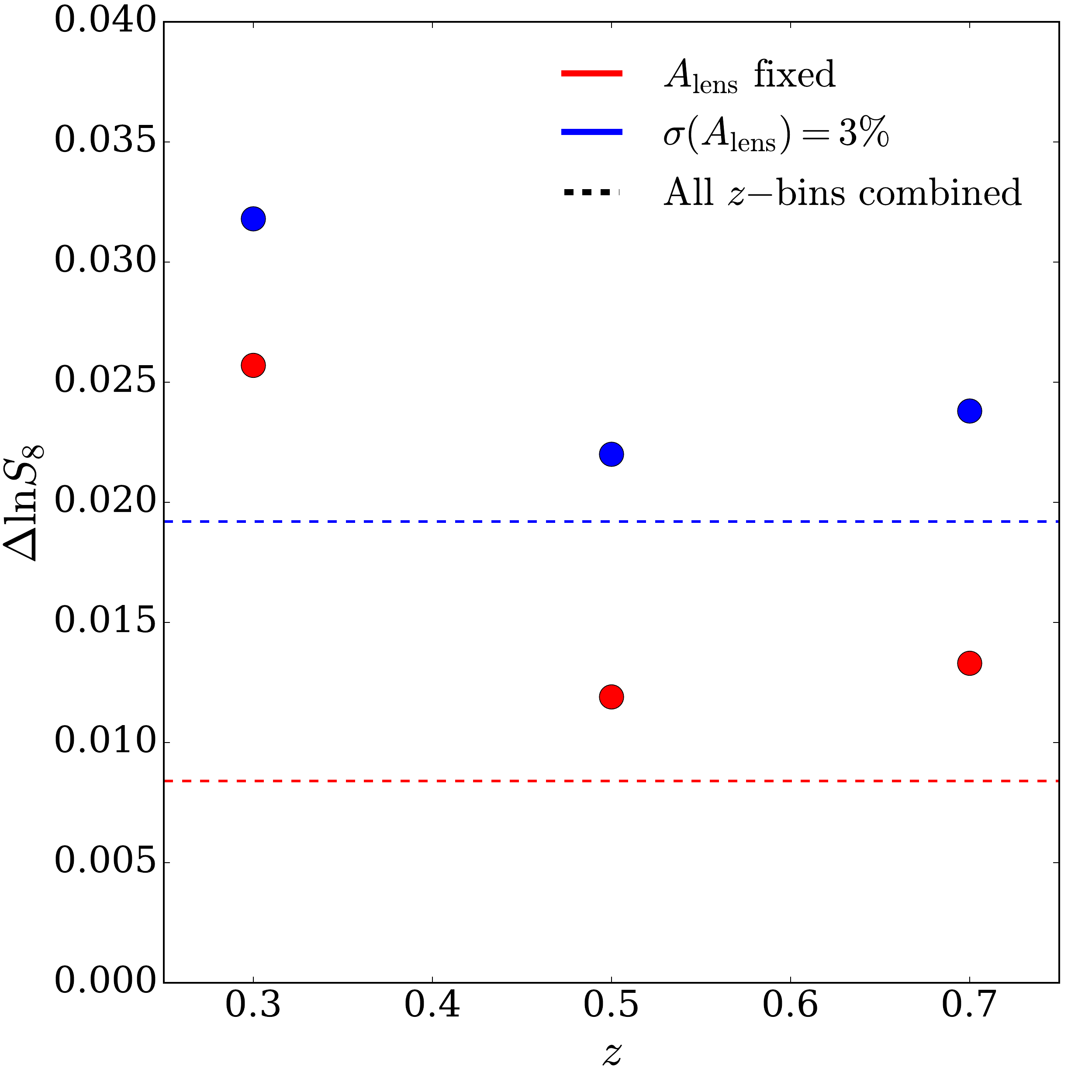}
\caption{Forecast constraints on $\ln S_8$ as a function of redshift marginalized over all HOD parameters with all other cosmological parameters fixed. Red points show constraints with $\Alens$ fixed while blue points show constraints from including our fiducial $3\%$ prior on $\Alens$. Analogously coloured dashed lines show constraints from combining all three of our redshift bins. When combining redshift bins we constrain HOD parameters in each bin separately.}
\label{fig:redshift}
\end{figure}

So far we have limited our forecasts to a bin of redshift $z = 0.35- 0.55$. Since DES redMaGiC galaxies extend from redshift $z = 0.15 - 0.70$, we now consider additional bins at lower and higher redshift. Specifically, we define three bins in redshift, $z = 0.15-0.35$, $z = 0.35 - 0.55$, and $z = 0.55-0.70$, and we use {\sc{abacus}} snapshots at $z = 0.3$, $z = 0.5$, and $z = 0.7$ respectively to compute emulator derivatives. We also compute separate covariance matrices for each bin taking into account the full range in redshift in each bin. Comparing across bins we observe little qualitative difference in derivatives for a given parameter. Quantitatively there is mild evolution, with most parameters having slightly larger effect at low redshift. A more important effect is the evolution of the covariance matrix. For $w_{p,gg}$ fractional errors decrease with increasing redshift because of increasing bin volume. For $\Delta \Sigma$ increasing volume with redshift is counteracted by fewer sources, which increases the shape noise contribution to the covariance and is dependent on the assumed source redshift distribution from \citet{Rozo_2011}. Fractional errors for $\Delta \Sigma$ improve going from the $z = 0.3$ to $z = 0.5$ redshift bin. However, going from $z = 0.5$ to $z = 0.7$ we find an increase in fractional error because the increase in volume is not able to compensate for the loss in sources.

Forecast results in all three redshift bins are shown in figure \ref{fig:redshift}. In each bin we forecast constraints on $S_8$ with our full datavector, $w_{p,gg}$ and $\Delta \Sigma$, with all other cosmological parameters fixed. For each bin we perform two separate forecasts in which $\Alens$ is fixed (red points) or free with a $3\%$ prior (blue points). Finally we indicate the constraint on $S_8$ from combining all three bins together with horizontal dashed lines. When combining constraints from multiple redshift bins we allow for different HOD parameters in each redshift bin and we assume redshift bins are independent. When $\Alens$ is free we forecast constraints of $3.2\%$, $2.2\%$ and $2.4\%$ on $S_8$ in the $z = 0.3$, $z = 0.5$ and $z = 0.7$ bins respectively. Fixing $\Alens$ improves these constraints to $2.6\%$, $1.2\%$ and $1.3\%$. As expected from our covariance matrices, we see that our constraint improves from $z = 0.3$ to $z = 0.5$. From $z = 0.5$ to $z = 0.7$ the constraint slightly degrades. In this case the precision has improved for $w_{p,gg}$ but gotten worse for $\Delta \Sigma$. When all three redshift bins are combined we forecast constraints of $1.9\%$ and $0.8\%$ on $S_8$ with $\Alens$ free and fixed, respectively. Both of these constraints slightly underperform simple quadrature combination of individual constraints.

\subsection{Summary}

\begin{figure*}
\centering
\includegraphics[width=1.0\textwidth]{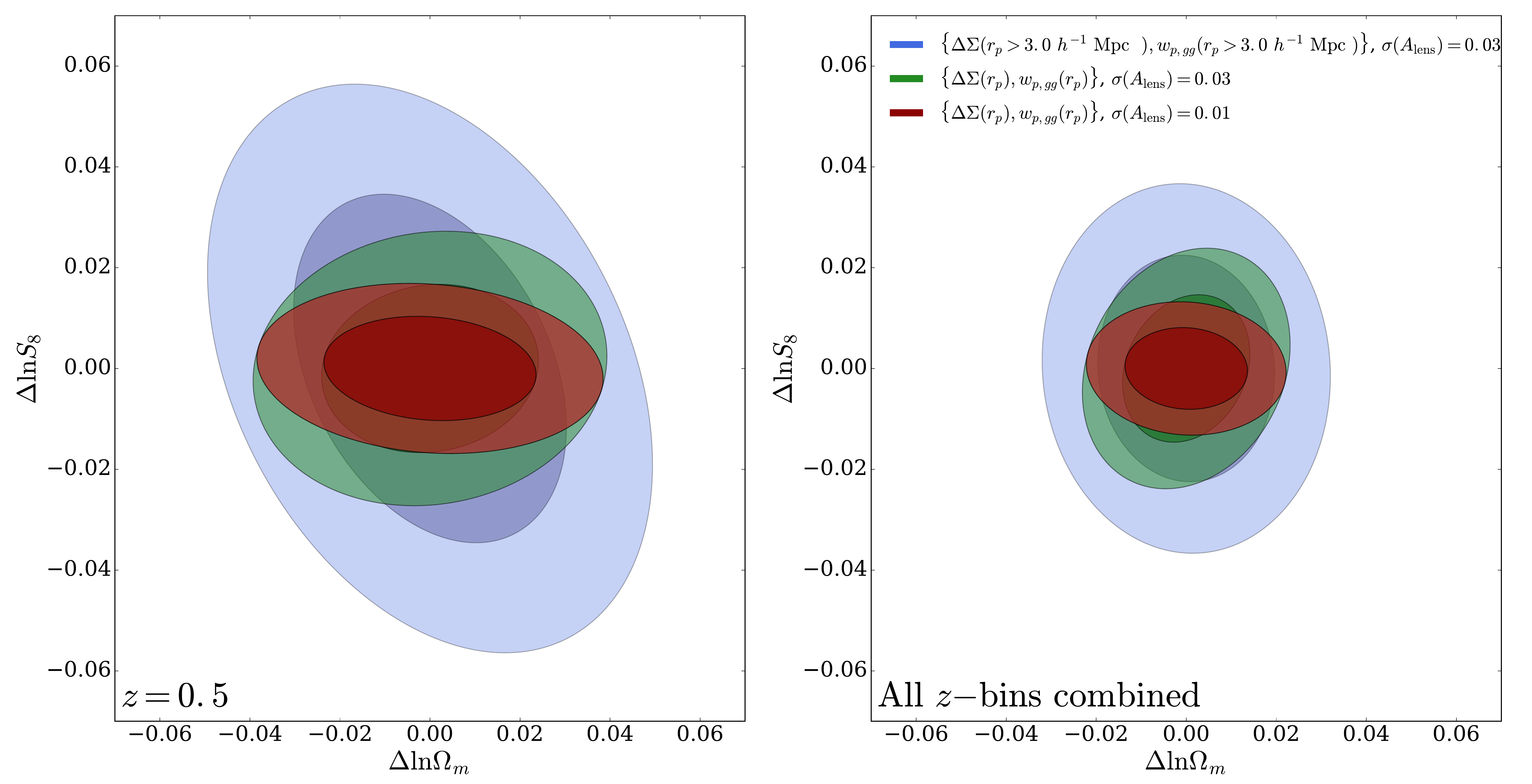}
\caption{Forecast constraints ($68\%$ and $95\%$ contours) on $S_8 = \sigma_8 \Omega_m^{0.5}$ and $\Omega_m$ with $H_0$, $n_s$, and $w_0$ fixed, summarizing some of our main results. Contours in the left panel show constraints from just the $z = 0.35{-}0.55$ bin while the right panel shows constraints from combining all three of our redshift bins. Blue contours show constraints when only the large scales of $\Delta \Sigma$ and $w_{p,gg}$ are used and a $3\%$ prior on $\Alens$ is assumed. Green contours show our fiducial scenario in which all scales of $\Delta \Sigma$ and $w_{p,gg}$ are used with a $3\%$ prior on $\Alens$. Red contours show constraints from all scales of $\Delta \Sigma$ and $w_{p,gg}$ when our prior on $\Alens$ is sharpened to $1\%$.}
\label{fig:S8-summary}
\end{figure*}

We have forecast cosmological parameter constraints for an analysis of galaxy-galaxy lensing $\Delta \Sigma$ and galaxy clustering $w_{p,gg}$ while marginalizing over a flexible HOD model and a scale independent lensing bias parameter $\Alens$. Figure \ref{fig:S8-summary} summarizes our main results in the $S_8{-}\Omega_m$ plane. The green contours in the left-hand panel show our fiducial scenario combining information from $\Delta \Sigma$ and $w_{p,gg}$ measured on scales $0.3 \, \Mpch < r_p < 30.0 \, \Mpch$ in a DES-like survey of galaxies within a bin of redshift $z=0.35{-}0.55$. For this scenario we forecast $3.2\%$ and $2.2\%$ constraints on $\Omega_m$ and $S_8$. When the `small' scales ($r_p < 3.0 \, \Mpch$) are omitted from such an analysis (blue contours) these constraints are degraded to $4.0\%$ and $4.6\%$ respectively. This difference in precision on $S_8$ is equivalent to a ${\sim}4.4$-fold increase in survey area, illustrating the stakes of accurate modeling of non-linear scales. If our external prior on $\Alens$ is be sharpened to $1\%$ (red contours) then constraints on $\Omega_m$ and $S_8$ sharpen even further to $3.1\%$ and $1.4\%$ respectively. 

In the right-hand panel of figure \ref{fig:S8-summary} we show results for the same three forecast scenarios when combining all three of our redshift bins spanning $z = 0.15 - 0.70$. When small scales of $\Delta \Sigma$ and $w_{p,gg}$ are omitted (blue contours) using all three bins of redshift we forecast a $2.6\%$ and $3.0\%$ constraint on $\Omega_m$ and $S_8$. This constraint on $S_8$ is an improvement on the $4.6\%$ constraint from the $z{=}0.5$ bin, but it is still relatively weak. When the small scales are also included in the datavector we forecast $1.9\%$ constraints on both $\Omega_m$ and $S_8$ (and $2.0\%$ on $\sigma_8$). These constraints are an improvement on the $3.2\%$ and $2.2\%$ obtained from the $z{=}0.5$ bin, though the $S_8$ gain is moderate in part because the $\Alens$ uncertainty affects all three redshift bins coherently. When the prior on $\Alens$ is reduced to $1\%$ these constraints improve to $1.8\%$ on $\Omega_m$, $1.4\%$ on $\sigma_8$, and $1.1\%$ constraint on $S_8$. This result shows the impressive gains that are attainable if future analyses can include small scale information from galaxy-galaxy lensing and clustering while controlling the uncertainty in lensing calibration over a broad range in redshift $z{=}0.15{-}0.70$. Our forecasts show that if those conditions are met the degeneracy between $\Omega_m$ and $\sigma_8$ can be broken to yield percent-level constraints on the amplitude of matter clustering.

\begin{figure}
\centering
\includegraphics[width=0.45\textwidth]{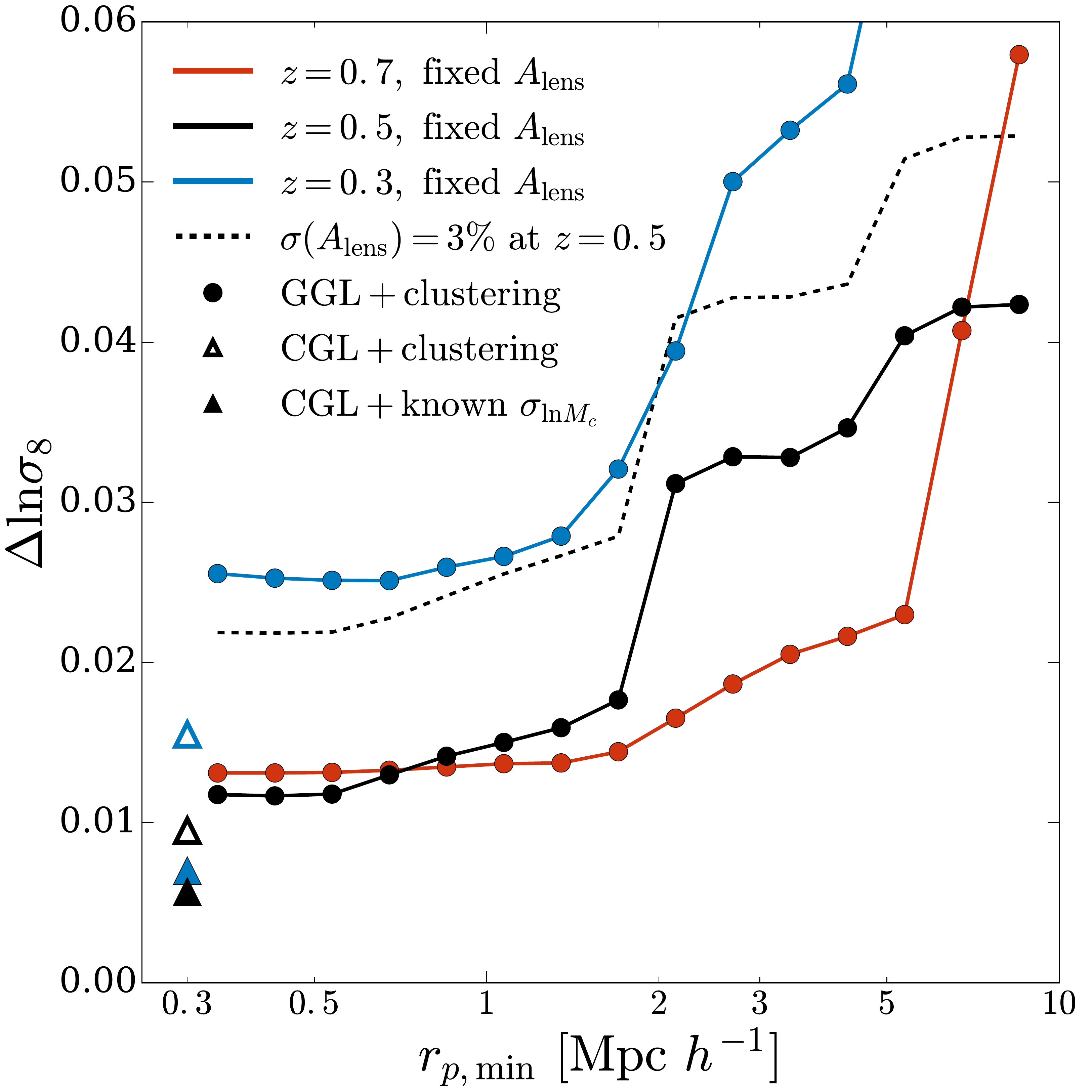}
\caption{Forecast constraints on $\ln \sigma_8$ as a function of minimum scale of $\Delta \Sigma$ and $w_{p,gg}$. The constraint on $\ln \sigma_8$ is marginalized over all HOD parameters with all other cosmological parameters fixed. The black solid line shows the case for the fiducial bin at $z{=}0.5$ with $\Alens$ fixed, while blue and red lines show results from the $z{=}0.3$ and $z{=}0.7$ bins. The dashed black line shows the case for the fiducial bin at $z{=}0.5$ with $\Alens$ included with our fiducial $3\%$ prior. For the sake of comparison analogously coloured triangles show constraints from the cluster weak-lensing $\Delta \Sigma$, $w_{p,cg}$ and $w_{p,gg}$ datavector of \citet{Salcedo_et_al_2020} using all scales of $0.3 < r_p < 30.0 \, \Mpch$. Filled in triangles show constraints for the case of fixed scatter in the cluster mass-observable relation $\siglnMc$, while empty triangles show the case of free $\siglnMc$.}
\label{fig:pfc}
\end{figure}

Figure~\ref{fig:pfc} presents a different summary form of our results, with an emphasis on the information content of smaller scales. Here we have forecast constraints on $\sigma_8$ with fixed values of the other cosmological parameters including $\Omega_m$; in a given redshift bin, fractional errors on $\sigma_8$ at fixed $\Omega_m$ are similar to the errors on $S_8$ with free $\Omega_m$.  Filled circles and connecting solid curves show forecast constraints for the three redshift bins with fixed $\Alens$, as a function of the minimum scale included in both $\Delta\Sigma$ and $w_{p,gg}$ (with $r_{p,\mathrm{max}}{=}30.0 \, \Mpch$ in all cases).  At $z{=}0.5$, the precision on $\sigma_8$ degrades moderately as $r_{p,\mathrm{min}}$ increases from $0.3\, \Mpch$ to $1.8\, \Mpch$, then degrades sharply as $r_{p,\mathrm{min}}$ crosses $2.0\, \Mpch$.  For $z{=}0.7$, the precision with small $r_{p,\mathrm{min}}$ is similar to $z{=}0.5$, and it degrades more slowly with increasing $r_{p,\mathrm{min}}$ until jumping sharply at $r_{p,\mathrm{min}}{=}8.0 \, \Mpch$. For $z{=}0.3$, the precision is lower as explained previously, and it is nearly constant for $r_{p,\mathrm{min}} \leq 1.0 \, \Mpch$.  The black dashed curve shows the forecast at $z{=}0.5$ with a $3\%$ prior on $\Alens$.  The $\Alens$ uncertainty significantly degrades the $\sigma_8$ precision, as shown previously in Table~\ref{table:margAlens-tests}, but the loss is smaller than one would expect from a naive quadrature combination of the $\Alens$ and  $\sigma_8$ uncertainties, even though both parameters have the same effect on $\Delta\Sigma$ in linear theory.  Determining $\sigma_8$ with a precision tighter than the $\Alens$ prior is a benefit of working into the non-linear regime, where the impact of $\sigma_8$ is scale-dependent.

Open triangles show the $\sigma_8$ precision forecasts from \citet{Salcedo_et_al_2020} for a combination of three observables: cluster weak lensing profiles $\Delta\Sigma_c(r_p)$, the projected cluster-galaxy cross-correlation function $w_{p,cg}(r_p)$, and the projected galaxy-galaxy correlation function $w_{p,gg}(r_p)$.  These forecasts are computed in the $z{=}0.15-0.35$ and $z{=}0.35-0.55$ redshift bins assuming DES-like cluster samples and weak lensing and clustering measurements, with fixed $\Alens$. We see that this three-observable combination can attain a $\sigma_8$ precision comparable to that of GGL+clustering at $z{=}0.5$ and better at $z{=}0.3$. \citet{Salcedo_et_al_2020} do not compute a forecast for $z{=}0.7$.  Although some systematics would be in common between these two analyses, such as uncertainties in shear calibration and source photometric redshifts, many systematics would be different. It is encouraging that clusters and GGL offer parallel routes to high-precision constraints on matter clustering from DES.  The three-observable combination considered by \citet{Salcedo_et_al_2020} constrains the scatter $\sigma_{\ln M_c}$ between true cluster mass and an observable mass proxy such as richness, which is the most important nuisance parameter that affects cosmological constraints from cluster weak lensing.  Filled triangles show the still tighter constraints that could be derived from cluster $\Delta\Sigma_c$ alone if $\sigma_{\ln M_c}$ were known independently.  \citet{Wu_et_al_2021} discuss cluster weak lensing constraints and the trade-off with $\sigma_{\ln M_c}$ and survey parameters in greater detail.

\section{Conclusions}
\label{sec:conc}

We have investigated potential cosmological constraints from a combination of galaxy-galaxy lensing $\Delta \Sigma$ and galaxy clustering $w_{p,gg}$ measured using redMaGiC selected galaxies with the precision expected in the final (Y6) data release of DES. We have computed observables using simulations from the {\sc{AbacusCosmos}} suite \citep{Garrison_et_al_2018} of N-body simulations and populating haloes with mock galaxies using a flexible HOD parameterization that includes central and satellite galaxy assembly bias. Using these observables we have constructed Gaussian process emulators \citep{Wibking_et_al_2020} of $w_{p,gg}$ and $\Delta \Sigma$, which accurately model each observable over a wide range of scales $0.3 - 30.0 \, \Mpch$ and a large space of HOD and cosmological parameters. We have also included in our analysis the effects of biased lensing calibration, represented by the parameter $\Alens$. We assume a fiducial HOD that is meant to describe the clustering of redMaGiC selected galaxies in DES; these values are listed in table \ref{table:params}. To compute covariance matrices we have used a mixture of analytic and numerical methods described in section \ref{sec:cov}. To represent potential measurements and modeling systematics, we have included a parameter $\Alens$ that multiplies all scales of $\Delta \Sigma$ by a common factor, and we have modified the weak lensing covariance matrix to analytically marginalize over a point mass contribution to $\Delta \Sigma$. These parameters can represent effects such as shear calibration bias, photo-z bias, or baryonic modification of halo density profiles on small scales.

With a $3\%$ prior on $\Alens$, we forecast precision of $1.9\%$ and $2.0\%$ on $\Omega_m$ and $\sigma_8$, respectively, from the combination of all three redshift bins, with fixed values of $H_0$, $n_s$, and $w_0$ and separate marginalization over all HOD parameters in each redshift bin.  The precision on $S_8$ is $1.9\%$.  If the prior on $\Alens$ is sharpened to $1\%$, then the $S_8$ constraint tightens to $1.1\%$.  Our results demonstrate the great promise of modeling GGL and galaxy clustering into the non-linear regime using HOD and N-body+emulator methods.  If we restrict our datavectors to scales $r_p \geq 3.0\, \Mpch$ then the $S_8$ precision degrades by a factor of 1.6, equivalent to a factor of 2.5 in survey area.  For the $1\%$ $\Alens$ prior the benefit of small scales is even larger, a factor of 2.8 in $S_8$ precision (a factor of 7.7 in equivalent survey area).  For the $z=0.5$ redshift bin, sections \ref{subsec:fid}-\ref{subsec:scales} examine the correlations between HOD and cosmological parameters, the impact of different systematics assumptions, and the contribution of different scales of the two observables (Figures~\ref{fig:variations}-\ref{fig:Alens} and Tables~\ref{table:margAlens-tests} and \ref{table:uncertainties}). In our forecasts, point-mass marginalization does not noticeably degrade cosmological parameter precision, but uncertainty $\geq 1\%$ in $\Alens$ does.
 
The recent DES-Y3 3$\times$2pt. cosmological analysis \citep{DES_3x2pt_2021} uses only large scale lensing and clustering data and obtains 9.3\%, 6.1\%, and 2.2\% constraints on $\Omega_m$, $\sigma_8$, and $S_8$. Comparison to our forecasts is difficult because this analysis includes cosmic shear, uses a magnitude-limited sample instead of redMaGiC, uses lower depth (Y3 vs.\ Y6) DES data, and includes nuisance parameters we have not considered here (such as intrinsic alignments). Closer to our scenarios, \cite{Pandey_DES_et_al_2021} analyzed DES-Y3 redMaGiC lensing and clustering in the linear regime, obtaining 10.7\% and 4.2\% constraints on $\Omega_m$ and $S_8$.  They caution that their $S_8$ results are likely biased by an unknown systematic causing internal inconsistency between redMaGiC lensing and clustering.  We have implicitly assumed that this challenge can be overcome by the time  of the final DES analyses and that remaining systematics can be adequately encapsulated by our $\Alens$ parameter even if they arise from multiple contributing factors.

Our emulator already appears accurate enough for the expected precision of final DES redMaGiC data (see Figure~\ref{fig:emulator}), though further testing and training on still larger simulation suites is desirable. We expect that our methods can be readily adapted to magnitude-limited samples, which should allow more precise $\Delta\Sigma$ measurements that require more careful treatment of photo-$z$ errors. Fortunately, in addition to affording high statistical precision, analyses that extend to non-linear scales provide rich opportunities for internal consistency checks and systematics tests, through distinctive scale dependence and comparison among galaxy samples that have different HODs but should yield consistent cosmological parameters.  For our $z{=}0.5$ forecast with all scales used in $w_{p,gg}$, we find essentially equal cosmological precision using scales $r_p > 3\, \Mpch$ and $r_p < 3\, \Mpch$ in $\Delta\Sigma$, allowing a strong consistency check between regimes where many systematics are very different.  If there is a 5-10\% discrepancy between low redshift matter clustering and CMB-normalized $\Lambda$CDM predictions, as suggested by some but not all recent weak lensing studies, then final DES analyses will demonstrate the discrepancy at high precision and allow initial explorations of its redshift and scale dependence.  Alternatively, if early universe fluctuations and low redshift matter clustering are consistent at the $1\%$ level, then maximally exploiting the potential of Stage III weak lensing surveys will demonstrate impressive success of standard cosmology and prepare the way for Stage IV dark energy experiments that are underway or beginning soon.

\section*{Acknowledgements}

We thank the creators of the Abacus Cosmos simulation suite for producing
this powerful public resource, with special thanks to Lehman Garrison and
Daniel Eisenstein for useful conversations.  We thank Ami Choi, Jack
Elvin-Poole, Niall MacCrann, Michael Troxel, and Ashley Ross for helpful
discussions of DES weak lensing and galaxy clustering. This work was supported in part by NSF grant AST-2009735. ANS was supported by a Department of Energy Computational Science Graduate Fellowship. This material is based upon work supported by the U.S. Department of Energy, Office of Science, Office of Advanced Scientific Computing Research, Department of Energy Computational Science Graduate Fellowship under Award Number DE-FG02-97ER25308. DW acknowledges the hospitality of the Institute for Advanced Study and the support of the W.M. Keck Foundation and the Hendricks Foundations during phases of this work. Simulations were analyzed in part on computational resources of the Ohio Supercomputer Center \citep{OhioSupercomputerCenter1987}, with resources supported in part by the Center for Cosmology and AstroParticle Physics at the Ohio State University. We gratefully acknowledge the use of the {\sc{matplotlib}} software package \citep{Hunter_2007} and the GNU Scientific library \citep{GSL_2009}. This research has made use of the SAO/NASA Astrophysics Data System.

This report was prepared as an account of work sponsored by an agency of the United States Government. Neither the United State Government
nor any agency thereof, nor any of their employees, makes any warranty, express or implied, or assumes any legal liability or responsibility
for the accuracy, completeness, or usefulness of any information, apparatus, product, or process disclosed, or represents that its use would
not infringe on privately owned rights. Reference herein to any specific commercial product, process, or service by trade name, trademark, manufacturer or otherwise does not necessarily constitute or imply its endorsement, recommendation, or favoring  by the United States Government or any agency thereof. The views and opinions of authors expressed herein do not necessarily state or relect those of the United State Government or any agency thereof.

\bibliographystyle{mnras}
\bibliography{masterbib}

\label{lastpage}

\end{document}